\documentclass[useAMS,usenatbib]{mn2e}
\usepackage{txfonts}
\usepackage{graphicx}
\usepackage{amssymb}
\usepackage[below]{placeins}
\usepackage{txfonts}
\usepackage{natbib}
\bibpunct{(}{)}{;}{a}{}{,}
\def \arcmin{$^{\prime}$}

\def \xmm{{\emph{XMM-Newton}}}

\def \chandra{{\emph{Chandra}}}

\def\vla{\emph{VLA}}
\def\gmrt{\emph{GMRT}}
\def \hst{{\emph{HST}}}

\def \galex{{\emph{Galex}}}

\newcommand{\ion}[2]{#1\,{\sc{#2}}}

\bibpunct{(}{)}{;}{a}{}{,}

\title[Violent interaction between the AGN and the hot gas in the core of the cluster S\'ersic 159-03]
{Violent interaction between the AGN and the hot gas in the core of the galaxy cluster S\'ersic 159-03}
\author[N. Werner et al.]{N. Werner$^1$\thanks{Chandra/Einstein fellow, E-mail: norbertw@stanford.edu}, M. Sun$^2$, J. Bagchi$^3$, S. W. Allen$^1$, G. B. Taylor$^4$\thanks{Adjunct Astronomer at the National Radio Astronomy Observatory}, S. K. Sirothia$^5$, A. Simionescu$^1$\thanks{Einstein fellow}, \newauthor   E. T. Million$^6$, J. Jacob$^7$, M. Donahue$^8$ \\
$^1$Kavli Institute for Particle Astrophysics and Cosmology, Stanford University, 452 Lomita Mall, Stanford, CA 94305-4085, USA \\
and SLAC National Accelerator Laboratory, 2575 Sand Hill Road, Menlo Park, CA 94025, USA \\ 
$^2$Department of Astronomy, University of Virginia, P.O. Box 400325, Charlottesville, VA 22904-4325, USA \\
$^3$Inter University Center for Astronomy and Astrophysics (IUCAA), Post Bag 4, Ganeshkhind, Pune 411 007, India \\
$^{4}$Department of Physics and Astronomy, University of New Mexico, Albuquerque, NM 87131, USA\\
$^5$National Centre for Radio Astrophysics, Tata Institute of Fundamental Research, Post Bag 3, Ganeshkhind, Pune 411007, India \\
$^6$Department of Physics and Astronomy, University of Alabama, Box 870324, Tuscaloosa, AL 35487, USA \\
$^7$Department of Physics, Newman College, Thodupuzha 685 585, India \\
$^8$Physics\&Astronomy Department, Michigan State University, East Lansing, MI 48824-2320, USA}

\begin{document}
\maketitle
\begin{abstract}
We present a multi-wavelength study of the energetic interaction between the central active galactic nucleus (AGN), the intra-cluster medium, and the optical emission line nebula in the galaxy cluster S\'ersic 159-03. We use X-ray data from \chandra, high resolution X-ray spectra and UV images from \xmm, H$\alpha$ images from the {\it SOAR} telescope, \hst\ optical imaging, and \vla\ and \gmrt\ radio data. The cluster center displays signs of powerful AGN feedback, which has cleared the central regions ($r<7.5$ kpc) of dense, X-ray emitting ICM. X-ray spectral maps reveal a high pressure ring surrounding the central AGN at a radius of $r\sim15$~kpc, indicating an AGN driven weak shock. The cluster harbors a bright, 44~kpc long H$\alpha$+[\ion{N}{ii}] filament extending from the centre of the cD galaxy to the north. Along the filament, we see low entropy, high metallicity, cooling X-ray gas. The gas in the filament has most likely been uplifted by `radio mode' AGN activity and subsequently stripped from the galaxy due to its relative southward motion. Because this X-ray gas has been removed from the direct influence of the AGN jets, part of it cools and forms stars as indicated by the observed dust lanes, molecular and ionized emission line nebulae, and the excess UV emission. 

\end{abstract}


\section{Introduction}

Supermassive black holes (SMBH) play a crucial role in galaxy formation. The correlation between the mass of the central SMBH and the bulge mass of their host galaxies suggests that their evolution is tightly coupled \citep{magorrian1998}. Outbursts of accreting SMBH (referred to as active galactic nuclei or AGN) disturb and heat the surrounding gas, lowering the accretion rate and, in some cases, driving so much gas out of the galaxy that the star formation is drastically reduced. 

The most dramatic portrayal of the interaction between AGN and their surroundings can be seen in the distribution of X-ray emitting gas in nearby giant ellipticals and clusters of galaxies. \chandra\ X-ray images have revealed giant cavities and shocks in the hot gas produced by repeated outbursts of the central AGN \citep[e.g.][]{fabian2003,nulsen2005,forman2005}. These AGN outbursts in principle provide enough power to offset radiative losses, suppress cooling, and prevent further star formation in the dense cores of clusters of galaxies \citep[see e.g.][]{mcnamara2007}. Detailed X-ray imaging and 2D spectral mapping of cooling cores with cavities, X-ray bright filaments, and shock fronts, provides the most reliable means of measuring the energy injected into the hot intra-cluster medium (ICM) by AGN, and is a powerful tool to study the physics of the AGN feedback in general \citep[see e. g.][]{randall2010,million2010,werner2010,simionescu2009b}. 

Here we present a detailed multi-wavelength study of AGN feedback processes in the core of the nearby \citep[z=0.0564;][]{maia1987}, relatively poor, low mass cluster of galaxies S\'ersic 159-03 (A S1101). 
The thermal properties and chemical enrichment of the hot ICM in this cooling core cluster have been studied in detail using \xmm\ \citep{kaastra2001,kaastra2004,deplaa2006}. Compared to other cooling core clusters, where the central temperature drops to $\sim$1/3 of the peak ambient value, S\'ersic~159-03 has a relatively modest (factor of 1.5) temperature drop in the core \citep{sun2009}. The radio properties of this cluster have been reported by \citet{birzan2008}, in the study of a sample of 24 cooling cores.
The bright optical emission line nebulae in the core of the cluster have been studied in detail by \citet{crawford1992} and more recently by \citet{oonk2010} using integral field spectroscopy. 

Throughout the paper we adopt a flat $\Lambda_{\mathrm{CDM}}$ cosmology with $H_{0}=70$ km$\, $s$^{-1}\, $Mpc$^{-1}$ and $\Omega_{M}=0.3$, which implies a linear scale of 1.2~kpc\, arcsec$^{-1}$ at the cluster redshift of $z=0.0564$ \citep{maia1987}.  Throughout the paper, abundances are given with respect to the solar values by \citet{grevesse1998}.
All errors are quoted at the 68 per cent confidence level.

\begin{table*}
\begin{center}
\caption{Summary of the radio observations. Columns list the central frequency, the telescope array, the resolution of the beam, its position angle, the rms noise, and the observing date.}
\begin{tabular}{cccccc}
\hline\hline
Frequency (MHz) & Array & Resolution (arcsec) & P.A. (deg) & rms Noise ($\mu$Jy beam$^{-1}$) & Obs. date \\
\hline 
244 & GMRT &  $20.04\times8.92$ & 11.0     & 1420  & 2007 November 2, 3 \\
325 & GMRT &  $17.67\times9.08$ & -179.4 & 352     & 2009 May 16,17  \\
617 & GMRT &  $8.29\times3.13$   & 9.7       & 250       & 2009 May 09,11  \\
1400 & VLA-B & $10\times4$          & 0.0        & 64       & 2006 August 8  \\
8400 & VLA-B & $3.1\times0.72$   & 8.3        & 21       & 2006 August 8  \\
\hline
\label{obs}
\end{tabular}
\end{center}
\end{table*}

\section{Data reduction and analysis}

\subsection{Optical, H$\alpha$, and UV data}

To study the optical properties of the central cluster galaxy, we analyzed a 600~s observation available in the {\it HST} archive (proposal ID: 8719). The image was taken with WFPC2 using the F555W filter and placing the galaxy onto the WF3 chip. 

Narrow-band optical imaging was performed with the 4.1 m {\it SOAR} telescope on September 13, 2009 (UT) using the {\it SOAR} Optical Imager (SOI).
The night was photometric with a seeing of around 1\arcsec. We used the 6916/78 CTIO narrow-band filter for the H$\alpha$+[\ion{N}{ii}] lines, and the 6738/50 filter for the continuum.
We took four 1200~s exposures with the 6916/78 filter and four 750~s exposures with the 6738/50 filter.
We reduced each image using the standard procedures in the {\tt IRAF MSCRED} package using EG~274 as a spectroscopic standard. The pixels were binned 2$\times2$, for a scale of 0.154\arcsec\ per pixel. More
details on the SOI data reduction can be found in \citet{sun2007}. 

We also analyzed near Ultra Violet data obtained by the \xmm\ Optical Monitor on 2002 November 20--21. Exposures were taken with the UVW1 (2400--3600 \AA) and UVW2 (1800--2400 \AA) filters.

\subsection{Radio data}

The summary of radio observations with \gmrt\footnote{The \gmrt\ is a national facility operated by the National Centre for Radio Astrophysics of the TIFR, India.} and \vla, including the observation date, obtained beam, and rms noise, is shown in Table~\ref{obs}. The strong, stable calibrator 3C\,48 was used for absolute flux density and bandpass calibration at all frequencies. We   tied  the final flux density scale to  the  standard `Baars-scale'  \citep{baars1977}. The VLA data were obtained in the B configuration. Their calibration was performed using {\tt AIPS} \citep{greisen2003} in the standard fashion, while imaging and self-calibration were performed using {\tt Difmap} \citep{shepherd1995}. 

The \gmrt\ data were reduced using {\tt AIPS++}~(version: 1.9). After applying bandpass corrections to the phase calibrator, gain and phase variations were estimated, and flux density, bandpass, gain and phase calibration were applied. The data for antennas with high errors in antenna-based solutions were examined and flagged over certain time ranges. Some baselines were also flagged based on closure errors on the bandpass calibrator. Channel and time-based flagging of data points corrupted by radio frequency interference (RFI) was applied using a median filter with a $6\sigma$ threshold. Residual errors above $5\sigma$ were also flagged after a few rounds of imaging and self calibration. The system temperature ($T_{\mathrm{sys}}$) was found to vary with antenna, the ambient temperature and elevation \citep{sirothia2009}. In the absence of regular $T_{\mathrm{sys}}$ measurements for \gmrt\ antennas, this correction was estimated from the residuals of corrected data with respect to the model. The corrections were then applied to the data. The final images were made after several rounds of phase self calibration, and one round of amplitude self calibration, where the data were normalized by the median gain for all the data. 

\subsection{\chandra\ X-ray data}

The \chandra\ observations of S\'ersic~159-03 were taken in August 2009 using the Advanced CCD Imaging Spectrometer (ACIS). The total net exposure time after cleaning is 89.6 ks. 
We follow the data reduction procedure described in \citet{million2009} and \citet{million2010}. 
Background spectra were extracted from the appropriate recent blank-sky fields available from the Chandra X-ray Center. These were normalized by the ratio of the observed and blank-sky count rates in the 9.5--12~keV band.

Background subtracted images were created in 13 narrow energy bands, spanning 0.5--7.5~keV. These were flat fielded with respect to the median energy for each image. The blank sky background fields were processed in an identical way to the source observation and reprojected to the same coordinate system. The background subtracted and exposure corrected images were co-added to create the broad band images. 

For the spectral analysis, point sources were excluded. 
The individual regions for the 2D spectral mapping were determined using the Contour Binning algorithm \citep{sanders2006b}, which groups neighboring pixels of similar surface brightness until a desired signal-to-noise threshold is met. In order to have small enough regions to resolve substructure and still have enough counts to achieve better than 7 per cent accuracy in the temperature determination, we adopted a signal-to-noise ratio of 35 ($\sim$1230 counts per region). Separate photon-weighted response matrices and effective area files were constructed for each region analyzed.

Spectral modeling has been performed with the {\tt SPEX} package \citep[{\tt SPEX} uses an updated version of the MEKAL plasma model with respect to {\tt XSPEC},][]{kaastra1996} in the 0.6--7.0 keV band.
To each bin we fitted a model consisting of absorbed \citep[Galactic $N_{\mathrm{H}}=1.14\times10^{20}$~cm$^{-2}$, ][]{kalberla2005} collisionally ionized equilibrium plasmas with temperature, spectral normalization (emission measure), and metallicity as free parameters.

\subsection{\xmm\ RGS data}

\begin{figure}
\includegraphics[width=0.85\columnwidth,clip=t,angle=0.,bb=48 36 565 757]{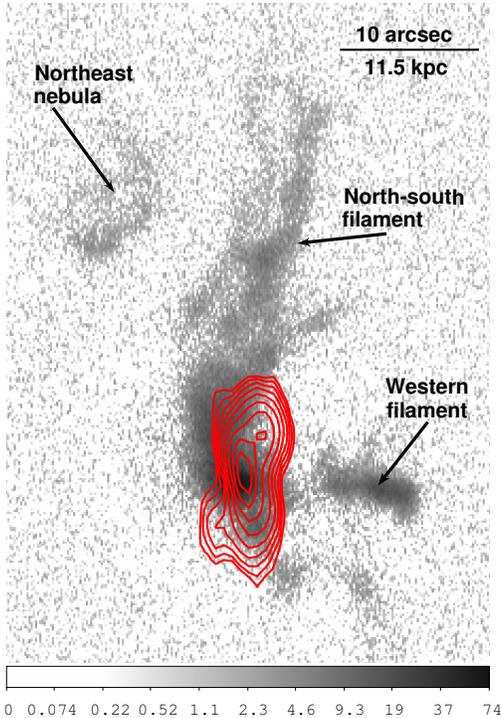}
\vspace{-0.2cm}
\caption{H$\alpha$+[\ion{N}{ii}] image obtained with the 4.1~m SOAR telescope. The image shows a large 44~kpc long north-south filament, a smaller western filament, and a separate nebula at the northeast. The northern end of the large filament separates into two parallel structures. The contours of the 8.4~GHz  radio emission are over-plotted in red. }
\label{Halpha}
\end{figure}

The hot plasma in the core of S\'ersic~159-03 exhibits a complex multi-temperature structure \citep{deplaa2006}. In order to determine the amount of low-temperature cooling X-ray gas in the core and obtain upper limits on the mass deposition rate, we analyze high spectral resolution \xmm\ Reflection Grating Spectrometer (RGS) X-ray data. 

The \xmm\ RGS data were obtained on November 20--21 2002, with a net exposure time of 86.3~ks. The data were processed as described in \citet{deplaa2006}. Spectra were extracted from a 4\arcmin\ wide extraction region, centred on the optical centre of the galaxy. Because the RGS operates without a slit, it collects all photons from within the 4\arcmin$\times\sim$12\arcmin\ field of view. Line photons originating at angle $\Delta\theta$ (in arcminutes) along the dispersion direction are shifted in wavelength by $\Delta\lambda=0.138\Delta\theta$~\AA. Therefore, every line is broadened by the spatial extent of the source. To account for this spatial 
broadening in our spectral model, we produce a predicted line spread function (LSF) by convolving the RGS response with the surface brightness profile of the cluster derived from the EPIC/MOS1 image along the 
dispersion direction. Because the radial profile of a particular spectral line can be different from the overall radial surface brightness profile, the line profile is multiplied by a scale factor $s$, which is the ratio of the 
observed LSF to the expected LSF. This scale factor is a free parameter in the spectral fit. We fit the spectra in the 8--28~\AA\ band.

\section{Results}

\begin{figure*}
\begin{minipage}{0.31\textwidth}
\includegraphics[width=5.7cm,clip=t,angle=0.,bb= 36 184 577 608]{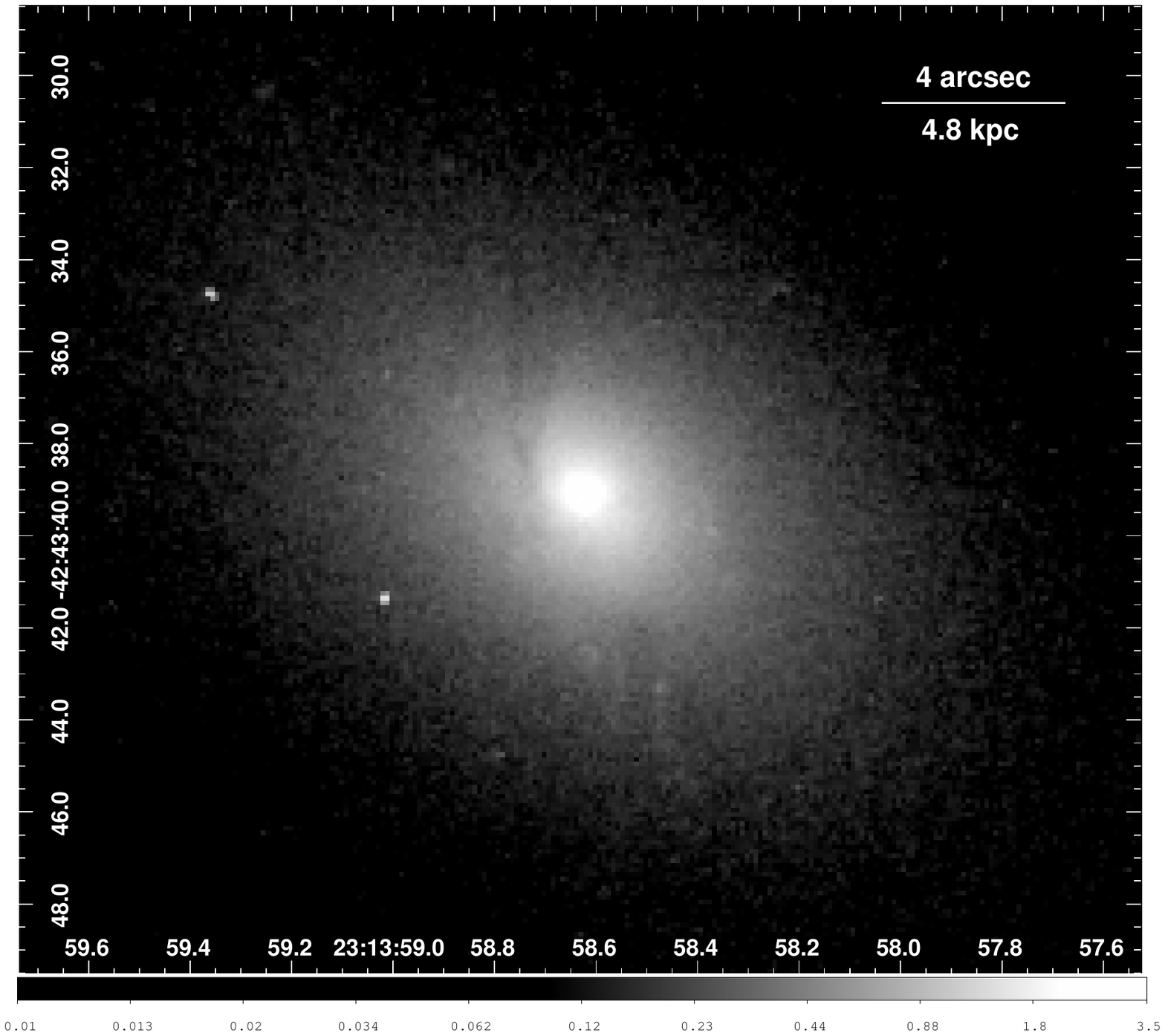}
\end{minipage}
\begin{minipage}{0.31\textwidth}
\includegraphics[height=5.2cm,clip=t,angle=0.,bb=36 175 577 617]{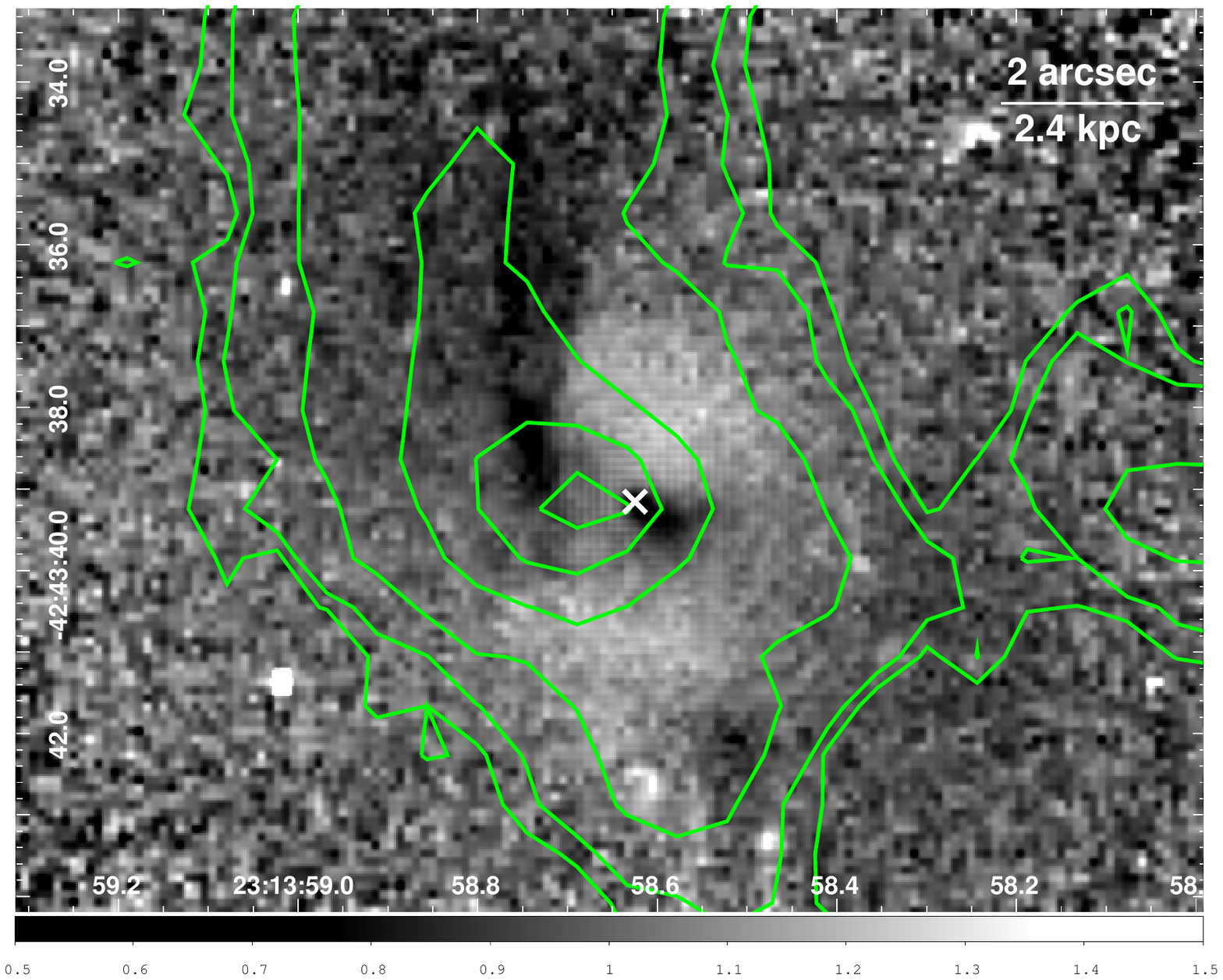}
\end{minipage}
\hspace{0.7cm}
\begin{minipage}{0.31\textwidth}
\includegraphics[width=5.cm,clip=t,angle=0.,bb= 36 118 577 674]{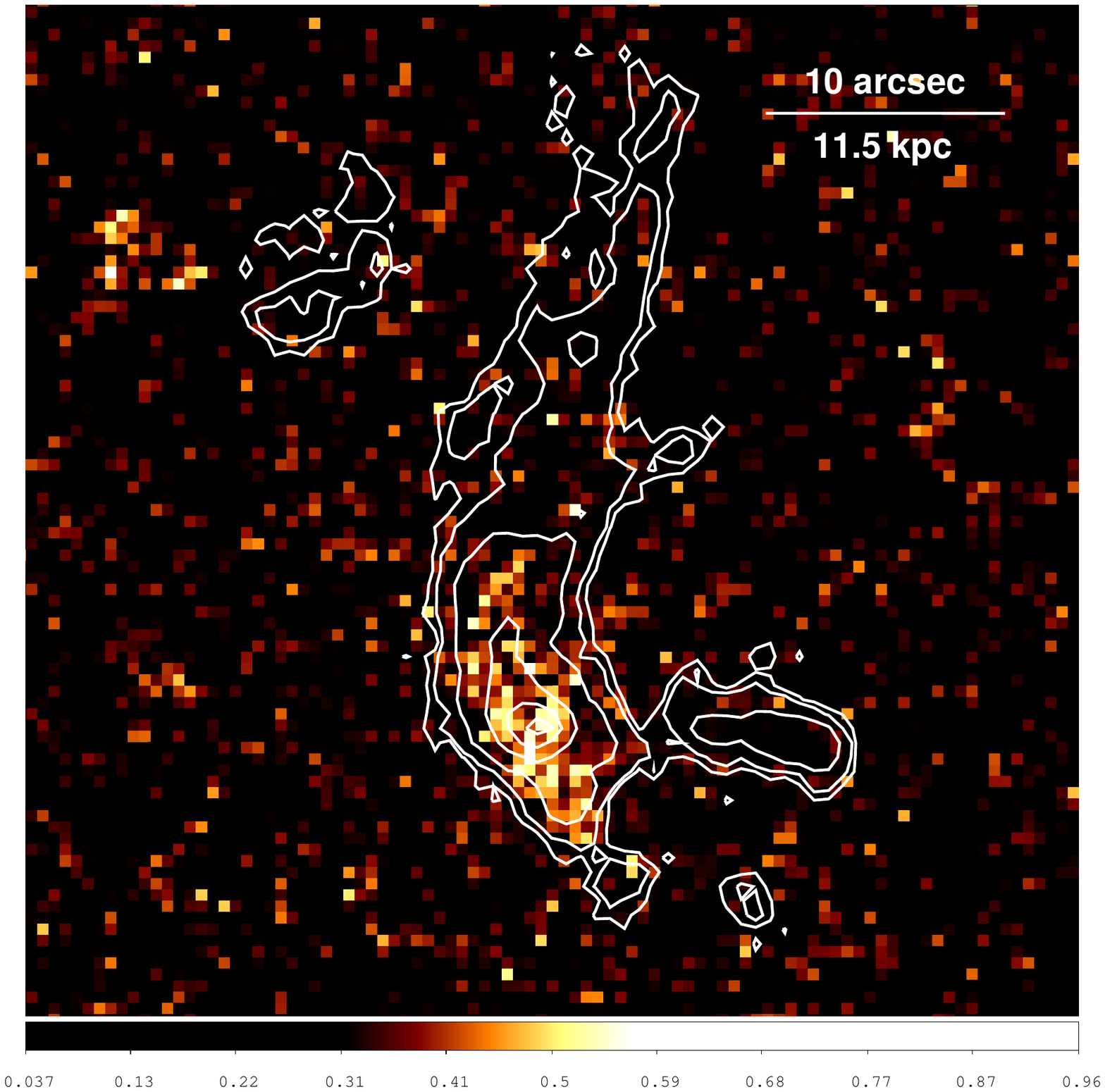}
\end{minipage}
\caption{{\it Left panel: }{\it HST} image of the central cluster galaxy of S\'ersic~159-03 with clear dark dust bands. {\it Middle panel: }The HST image divided by the best fit 2D elliptical de Vaucouleurs model of the galaxy. The white ``X'' mark indicates the optical center of the galaxy. A 4.5~kpc long dust lane displaced to the east and extending to the north is clearly visible. The inner region of the galaxy is disturbed and not well described by the de Vaucouleurs profile. The green contours show the isophotes of the H$\alpha$+[\ion{N}{ii}] filaments (see Fig.~\ref{Halpha}). {\it Right panel: } \xmm\ Optical Monitor UVW2 (1800--2400 \AA) image of of the core of the galaxy with the H$\alpha$ contours over-plotted. The UV emission clearly extends along the brightest region of the H$\alpha$ filament. }
\label{HST}
\end{figure*}

\subsection{Optical and UV properties}

The cooling core of S\'ersic~159-03 harbors a remarkable H$\alpha$+[\ion{N}{ii}] emission line filament system shown in Fig.~\ref{Halpha}. This system of emission line nebulae consists of a large 44~kpc long north-south filament extending to the radius of 35~kpc in the north and a smaller western structure extending out to 16~kpc. The total H$\alpha$+[\ion{N}{ii}] flux is $5.9\times10^{-14}$ erg~s$^{-1}$~cm$^{-2}$.  Assuming [\ion{N}{ii}]$_{6583\AA}$/H$\alpha$ = 1 and [\ion{N}{ii}]$_{6548\AA}$/[\ion{N}{ii}]$_{6583\AA}$ = 0.35, the total H$\alpha$ flux is $\sim$$2.5\times10^{-14}$ erg~s$^{-1}$~cm$^{-2}$.  The total luminosity of the H$\alpha$ line emission (with the [\ion{N}{ii}]  flux excluded) is $L_{\mathrm{H\alpha}}=2\times10^{41}$~ergs~s$^{-1}$.  
Our measured H$\alpha$ flux is higher than the flux $f_{\mathrm{H}\alpha}=1.79\pm0.15\times10^{-14}$ erg~s$^{-1}$~cm$^{-2}$ measured by \citet{mcdonald2010}, who used a filter with a 10 times narrower bandpass targeting only the H$\alpha$ emission lines. 
A separate nebula can be seen 25~kpc to the northeast of the core. 

Assuming a volume filling fraction of unity, a cylindrical geometry in the H$\alpha$ bright regions, optically thin ionized gas, isotropic radiation, and case B recombination at 10$^4$~K \citep{bhm1999}, the total mass of the H$\alpha$ emitting gas is about $2.7\times10^9$~$M_\odot$. However, these optical filaments most likely consist of many thin threads with very small volume filling fractions \citep[e.g.][]{fabian2008}. The inferred mass is therefore likely to be significantly overestimated. 

The giant elliptical galaxy in the core of S\'ersic~159-03 has its major axis aligned in the northeast-southwest direction (left panel of Fig.~\ref{HST}). The HST data reveal a 4.5~kpc long dust lane extending from the centre to the north, coincident with the bright H$\alpha$ emitting gas. It is displaced to the east from the core of the galaxy by about 1.2\arcsec\ ($\sim$1.4~kpc). The dust lane is clearly visible in the middle panel of Fig.~\ref{HST}, where the HST image is divided by the best fit 2D elliptical de Vaucouleurs model. This ratio image also uncovers a fainter dust lane to the south of the core. 
The over-subtraction of the emission from center of the galaxy by the model shows that the surface brightness of the stellar core is flatter than the de Vaucouleurs profile. We do not detect optical emission from the AGN. 

Near Ultra Violet images (NUV) from the \xmm\ Optical Monitor show diffuse emission which is co-spatial with the dust lanes and the emission line nebulae, providing a strong indication for current star-formation (right panel of Fig.~\ref{HST}). 
After correcting for coincidence, dead-time loss, and time sensitivity degradation, the total UVW1 luminosity within a radius of 7~arcsec is 3.7$\times10^{42}$ ergs s$^{-1}$. Based on \citet{cardelli1989}, the Galactic extinction in the UVW1 band was assumed to be 0.07~mag and the unknown internal extinction was set to zero. We measured the 2MASS J-band luminosity within the same aperture and subtracted the contribution from the passive stellar population from the empirical $L_{\mathrm{UVW1}} - L_{\mathrm{J}}$ relation derived by \citet{hicks2005}. We find a net NUV luminosity excess of $\sim$2.3$\times10^{42}$ ergs s$^{-1}$. Without correcting for internal extinction, we obtain a UVW2-UVW1 color $\sim$0.4 mag.

\subsection{Radio morphology}

\begin{figure*}
\hspace{-2cm}
\begin{minipage}{0.32\textwidth}
\includegraphics[width=7.0cm,clip=t,angle=0.]{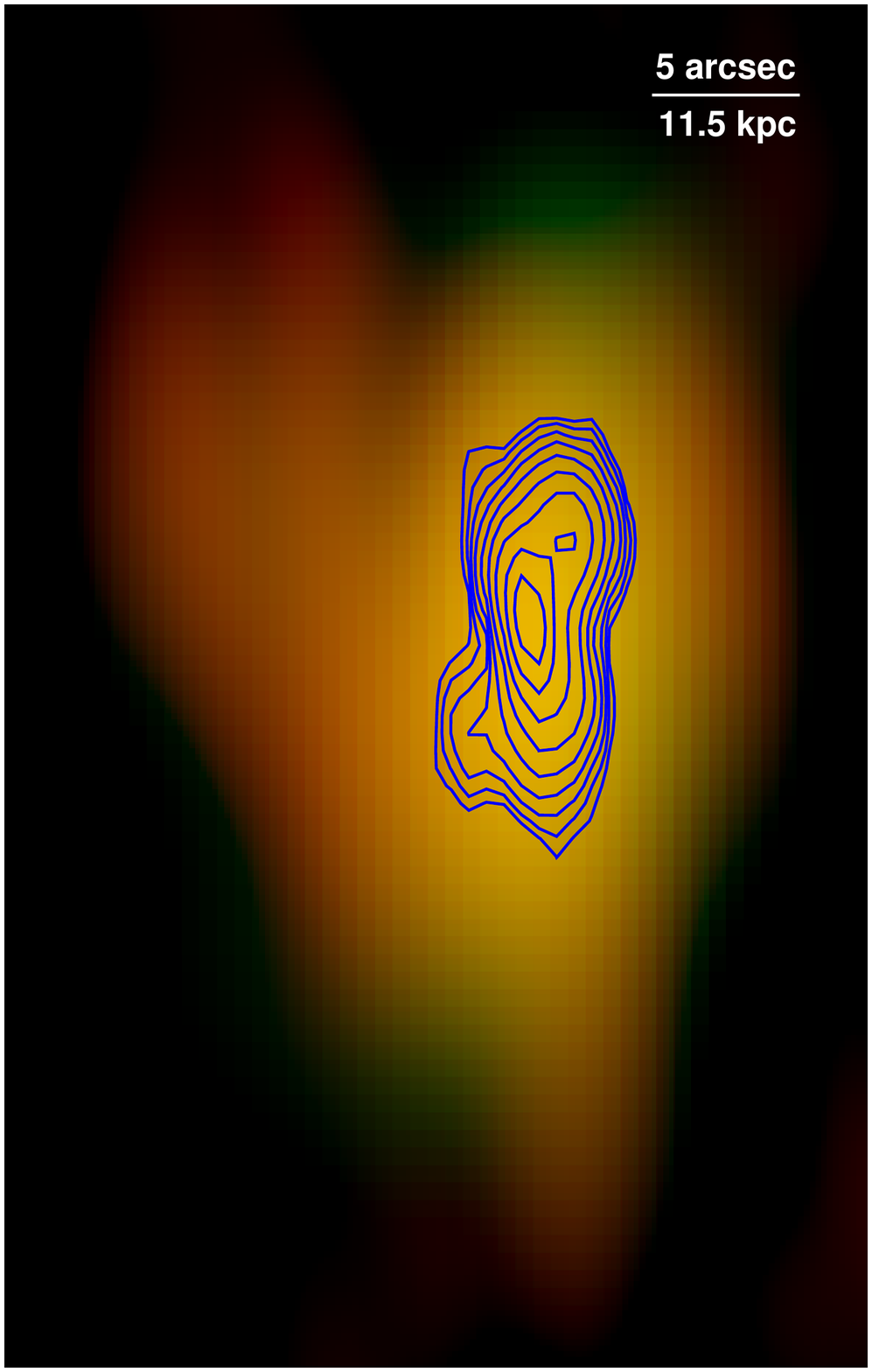}
\end{minipage}
\hspace{0cm}
\begin{minipage}{0.32\textwidth}
\includegraphics[width=7.0cm,clip=t,angle=0.]{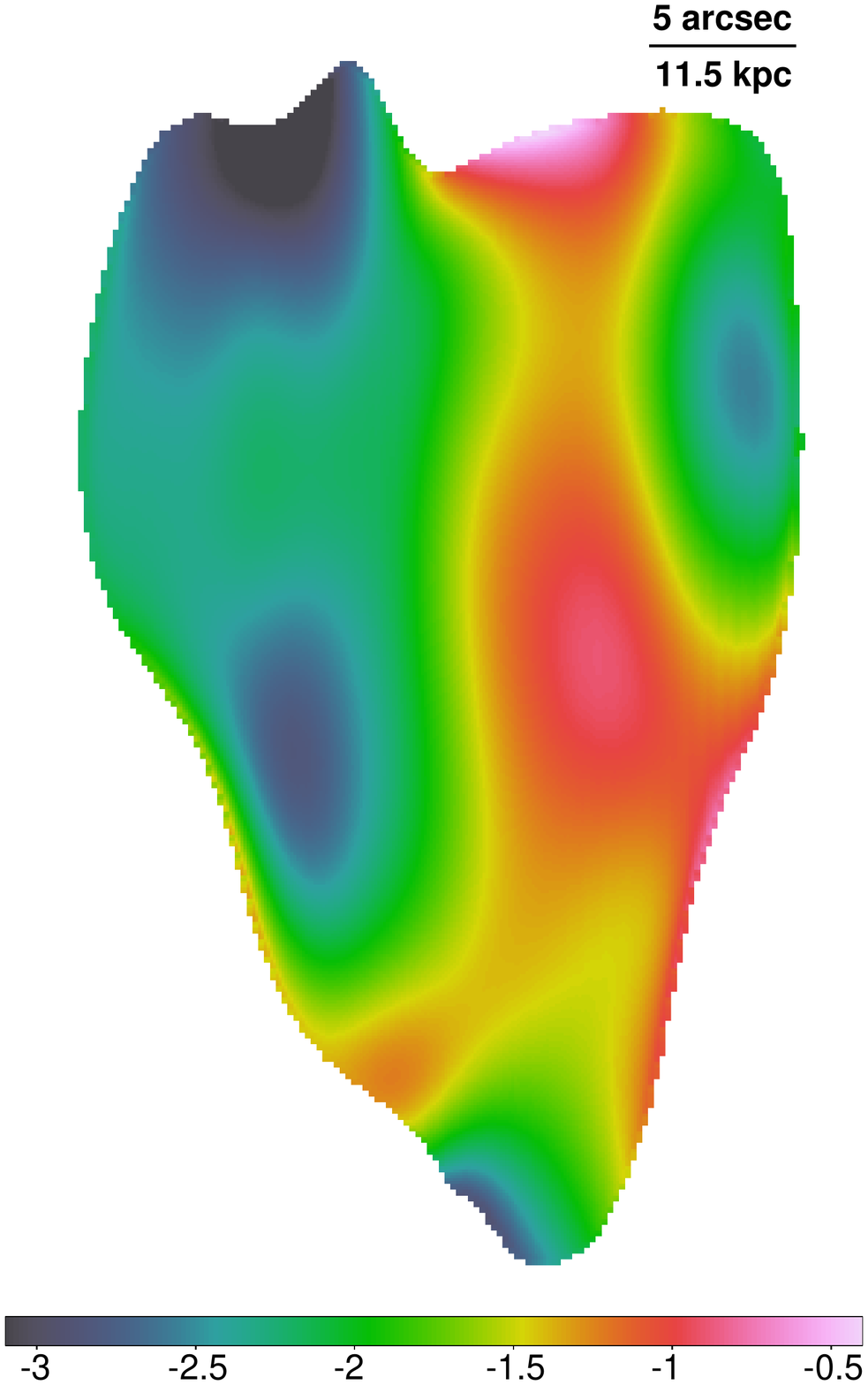}
\end{minipage}
\hspace{0cm}
\begin{minipage}{0.32\textwidth}
\includegraphics[width=8cm,clip=t,angle=0.]{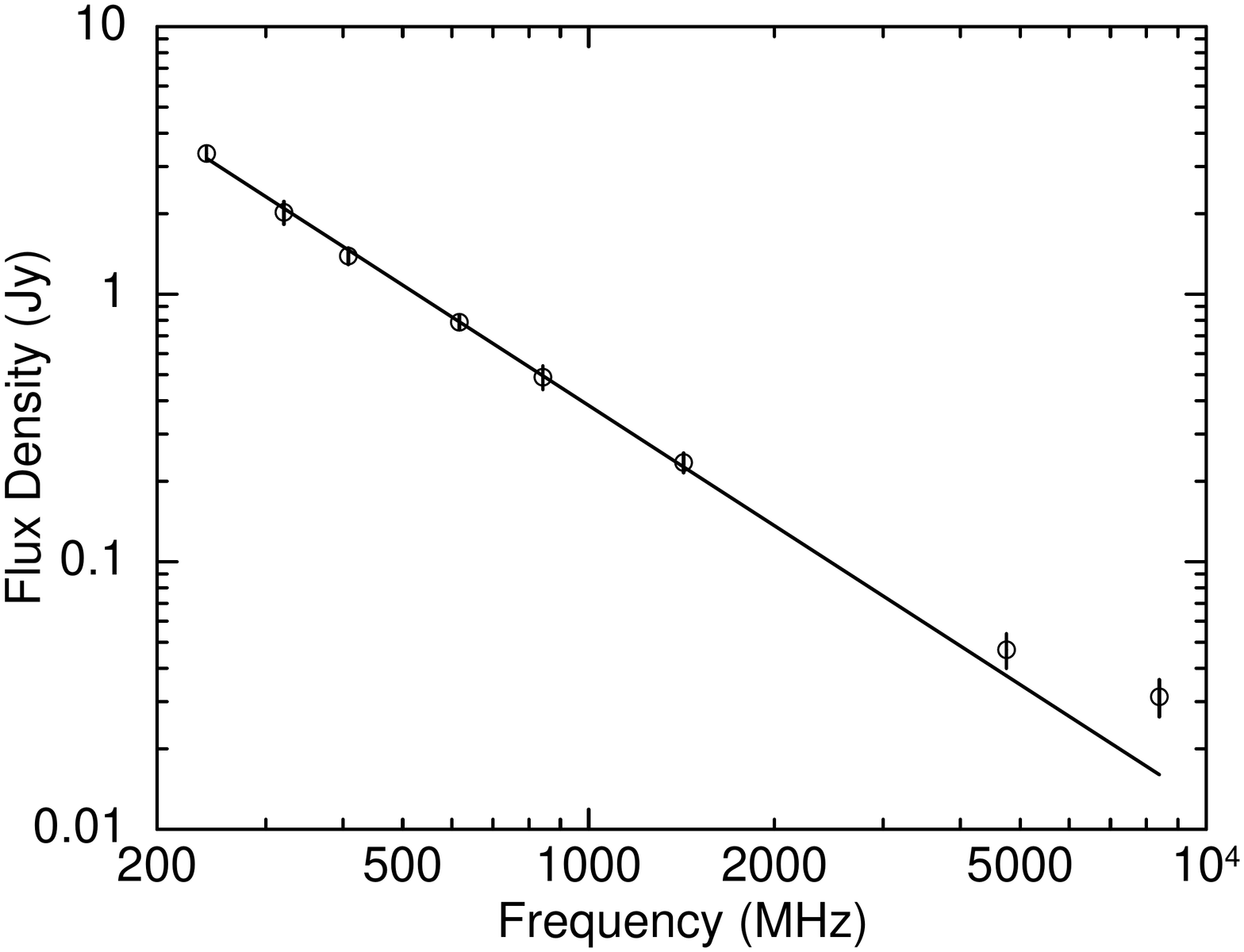}
\end{minipage}
\caption{{\it Left panel:} Red+green color image, produced from the radio data at 617~MHz (red) and 1.4~GHz (green), with the 8.4 GHz radio contours overplotted. {\it Central panel:} Spectral index map produced by combining the 617~MHz \gmrt\ and the 1.4~GHz \vla\ radio data. The radio maps show the innermost 0.5$\times$0.78 arcmin$^2$ region of the cluster. The radio maps at the two frequencies have been matched in resolution.  {\it Right panel:} The broad band radio spectrum between 244~MHz and 8.4~GHz. The radio spectrum between 244~MHz and 1.4~GHz can be described by the overplotted power-law $S_{\nu} \propto \nu^{-\alpha}$ with index $\alpha=1.49$.}
\label{radiospec}
\end{figure*}

\begin{table}
\begin{center}
\caption{Radio flux densities. The flux densities obtained by GMRT and VLA are from this work, the values obtained at 408 MHz and 843 MHz by MOST are from the MRC \citep{large1981} and SUMSS surveys \citep{bock1999,mauch2003}. The flux density at 4.75~GHz obtained by {\it ATCA} is from Hogan et al. in preparation.}
\begin{tabular}{lcc}
\hline\hline
Frequency & Array & Flux density \\
(MHz) & &  (Jy) \\
\hline 
244 & GMRT &  $3.360\pm0.20$ \\
325 & GMRT &  $2.024\pm0.20$ \\
408 & MOST & $1.390\pm0.10$ \\
617 & GMRT & $0.786\pm0.05$\\
843 & MOST & $0.472\pm0.014$ \\
1425 & VLA-B & $0.235\pm0.020$ \\
4752 & ATCA & $0.0469\pm0.007$\\
8400 & VLA-B & $ 0.0313\pm0.005$\\
\hline
\label{radiofluxes}
\end{tabular}
\end{center}
\end{table}

The contours of the 8.4~GHz radio emission in Fig.~\ref{Halpha} and \ref{radiospec} resolve the source into an S-shape, oriented primarily north-south, with a compact central feature of flux density 6.3 mJy/beam, which is approximately the fifth of the total radio flux at this frequency. The jets seem to start out in the northeast-southwest direction, before bending clockwise. At 1.4 GHz the total flux density is $\sim$230 mJy and shows a core with an extension $\sim$12 arcseconds to the east (see the white contours in Fig.~\ref{images}). The centers of the radio and optical emission are well aligned. 
At the 617~MHz \gmrt\ radio map, the morphology of the radio emission is nearly identical to 1.4~GHz. The extension to the east is robustly detected at 617~MHz and seems to be divided into two filaments. The lower frequency GMRT radio maps do not significantly resolve the radio source. 

In the left panel of Fig.~\ref{radiospec} we show a red+green color image, produced from the radio data at 617~MHz (red) and 1.4~GHz (green) with the contours of the 8.4~GHz radio emission overplotted. 
This combined radio image clearly shows the extension to the east.  The central panel of Fig.~\ref{radiospec} shows the spectral index map of the core of S\'ersic~159-03 produced by combining the 617~MHz \gmrt\ and the 1.4~GHz \vla\ radio data. Only data above 2 mJy/beam at 617 MHz and 0.2 mJy/beam at 1.4 GHz were included. To the east, north-east and north-west of the core the radio emission has a very steep spectrum described by a power-law $S_{\nu} \propto \nu^{-\alpha}$ with index $\alpha>2.2$. The spectrum of the central region is flatter with index $\alpha\sim1$. 

In Table~\ref{radiofluxes} we show the integrated radio flux densities at five different frequencies obtained with \gmrt\ and \vla. We also list the published flux densities obtained by the MRC survey at 408~MHz \citep{large1981} and by the SUMSS survey at 843~MHz \citep{bock1999,mauch2003} performed with the Molonglo Observatory Synthesis Telescope (MOST). The integrated flux density at 4.75 GHz obtained by the Australia Telescope Compact Array (ATCA) is from Hogan et al. in preparation. The broad band radio spectrum is shown in the right panel of Fig.~\ref{radiospec}. Between 244~MHz and 1.4~GHz the integrated spectrum can be described by a power-law $S_{\nu} \propto \nu^{-\alpha}$ with $\alpha=1.49$. At high frequencies, between 4.8~GHz and 8.4~GHz, however, the best fit power-law is significantly flatter with index $\alpha=0.71$.

\subsection{X-ray imaging: a disturbed morphology}

\begin{figure*}
\begin{minipage}{0.47\textwidth}
\includegraphics[height=8.3cm,clip=t,angle=0.,bb= 36 135 577 657]{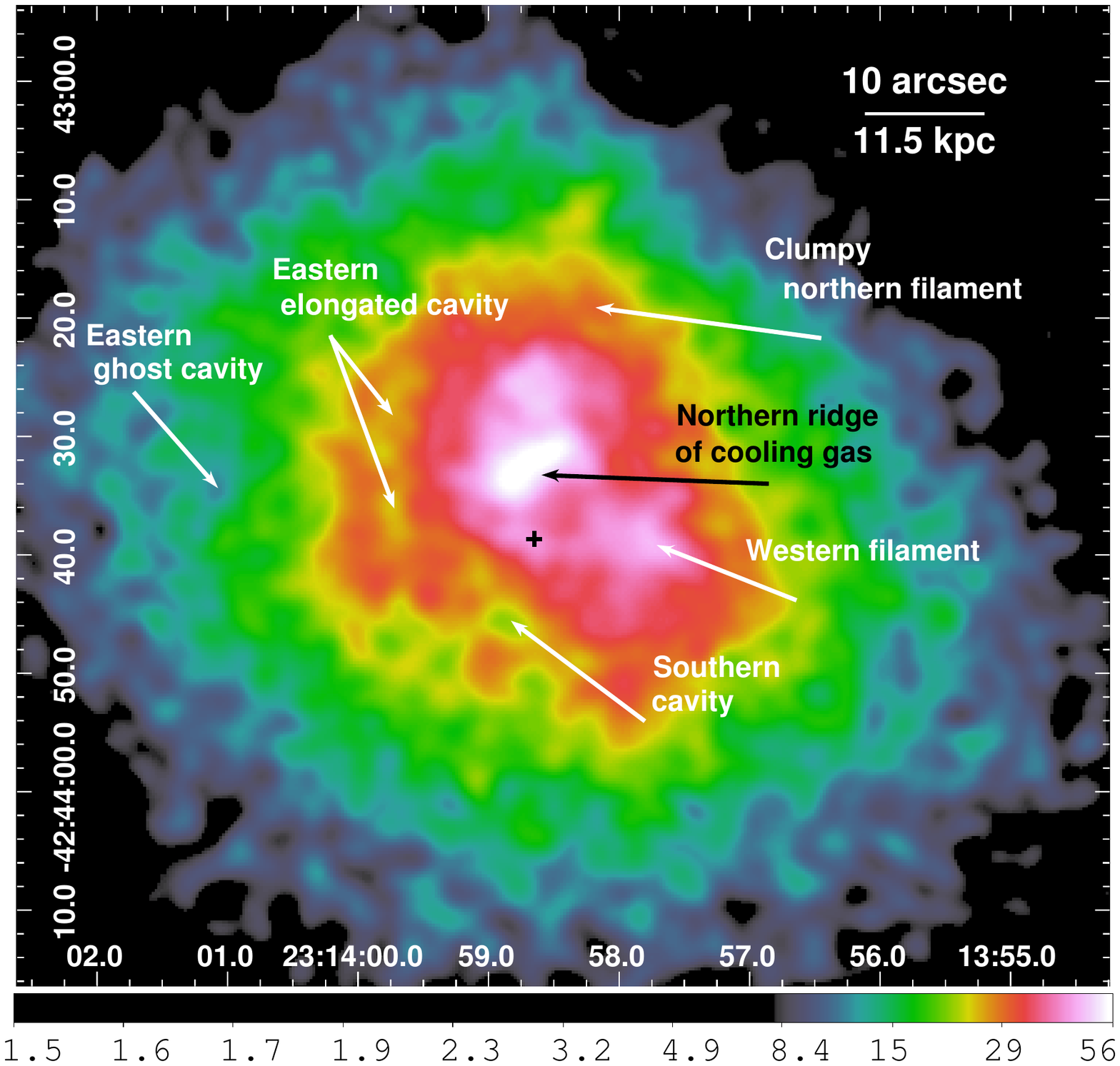}
\end{minipage}
\begin{minipage}{0.47\textwidth}
\includegraphics[height=8.3cm,clip=t,angle=0.,bb= 36 135 577 657]{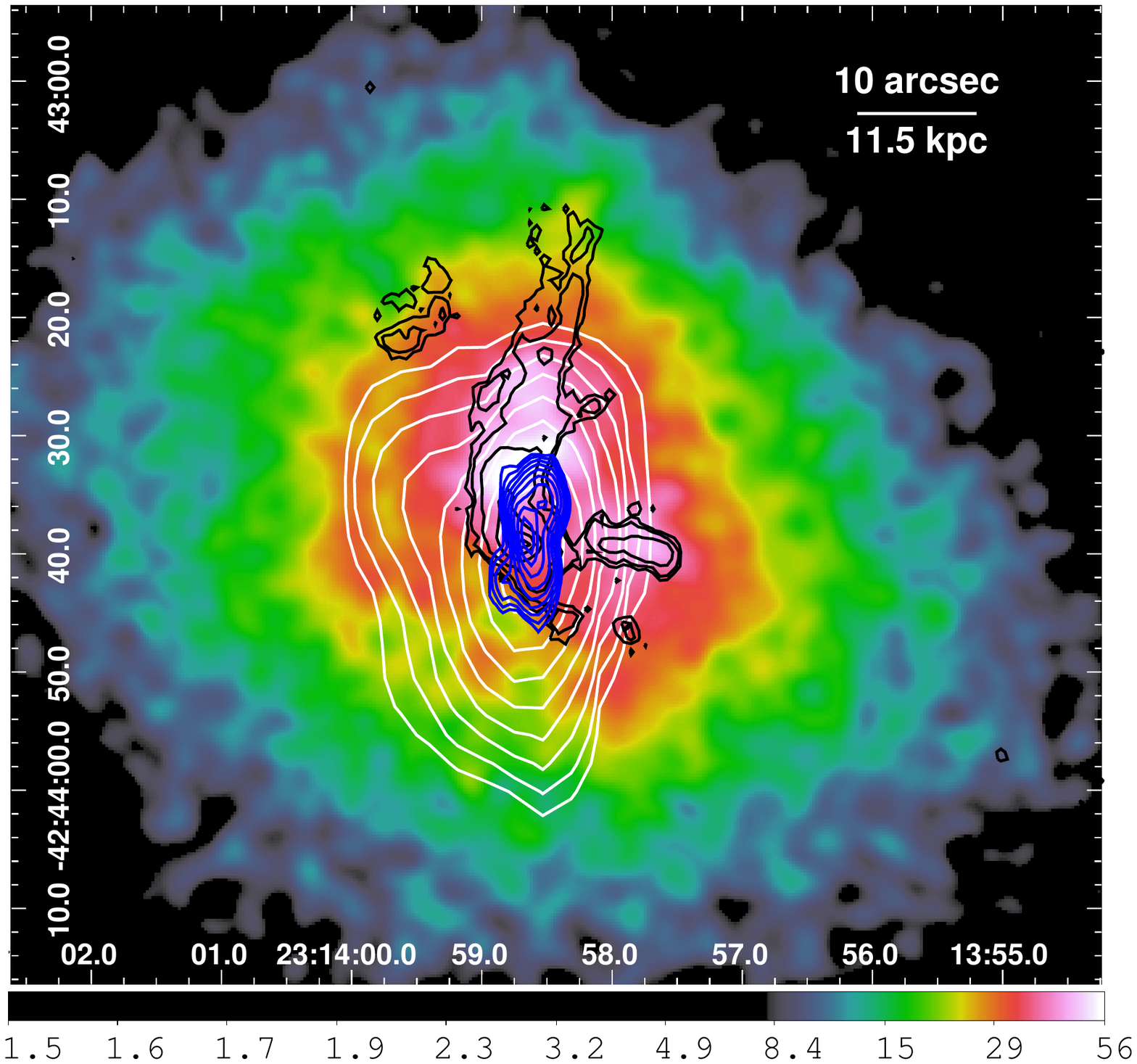}
\end{minipage}\\
\vspace{2mm}
\begin{minipage}{0.47\textwidth}
\includegraphics[height=8.3cm,clip=t,angle=0.,bb= 36 135 577 657]{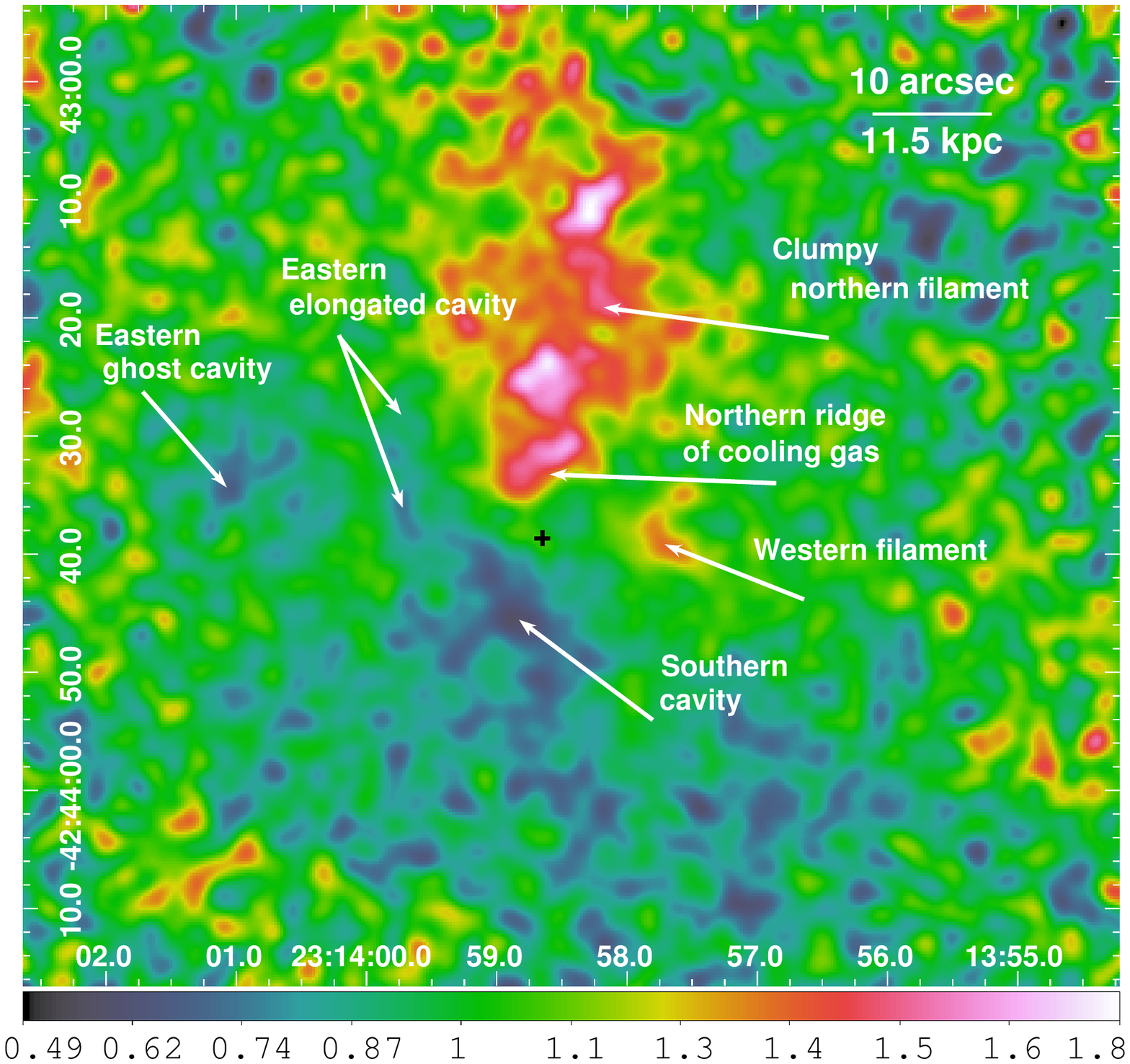}
\end{minipage}
\begin{minipage}{0.47\textwidth}
\includegraphics[height=8.3cm,clip=t,angle=0.,bb= 36 135 577 657]{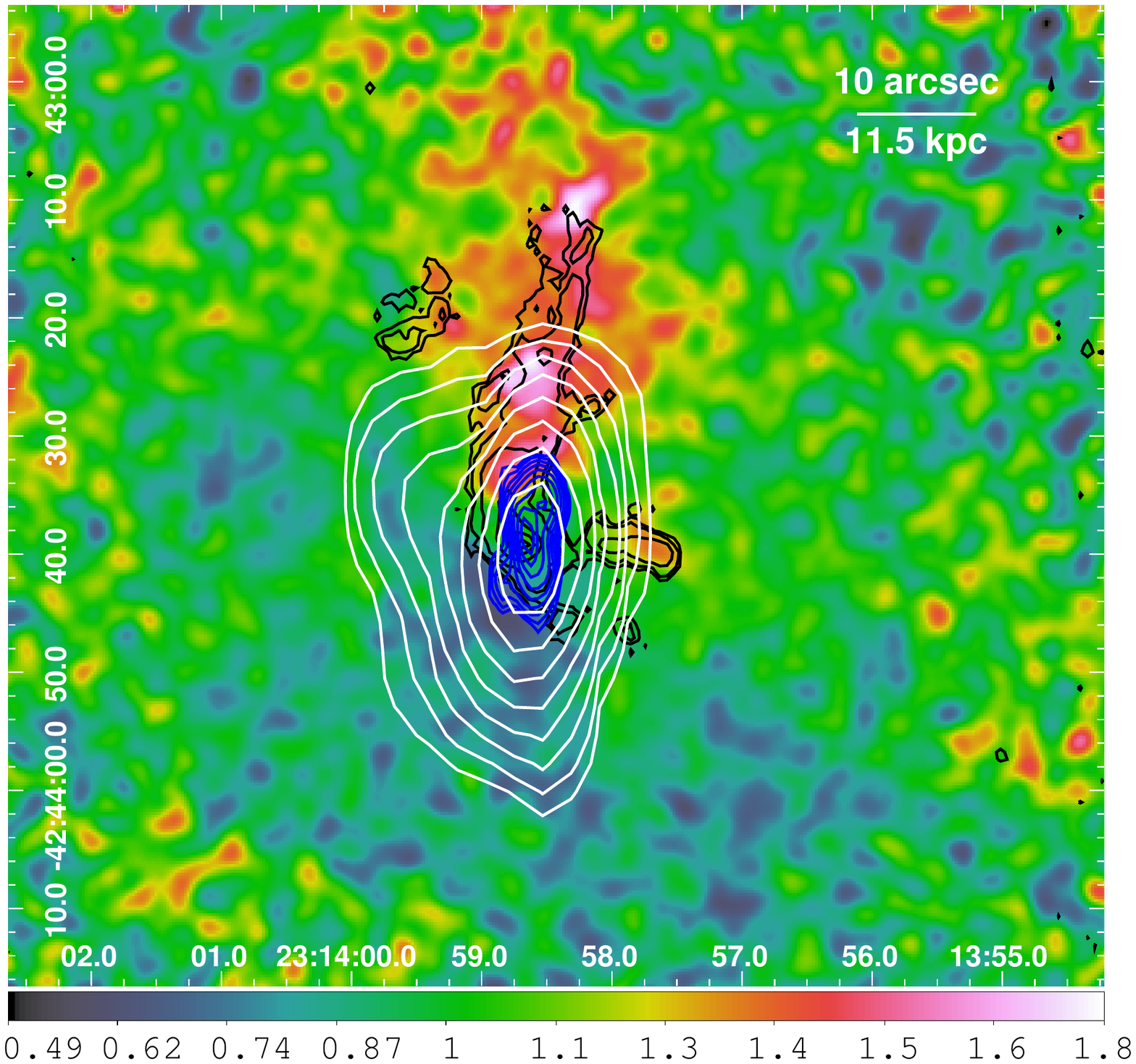}
\end{minipage}
\caption{{\it Top left panel:} Background subtracted, flat-fielded \chandra\ X-ray image of the core of S\'ersic~159-03 in the 0.5--7.5~keV band. The cross indicates the position of the central radio source. {\it Top right panel:}  The same \chandra\ image with over-plotted contours of the 8.4~GHz (blue) and 1.4~GHz (white) radio emission, and with the contours of the H$\alpha$+[\ion{N}{ii}] line emission (black). The dense cooling gas is displaced to the north of the centre of the galaxy and the region within $r\lesssim7.5$~kpc is void of dense gas. {\it Lower left panel:} The \chandra\ image divided by the best fit 2D elliptical double beta model. Note the clear structure to the north.  {\it Lower right panel:} The same ratio image with the contours of the radio and H$\alpha$+[\ion{N}{ii}] emission over-plotted. The images have been smoothed with a Gaussian kernel.  }
\label{images}
\end{figure*}

\begin{figure*}
\begin{minipage}{0.47\textwidth}
\includegraphics[height=8.3cm,clip=t,angle=0.,bb= 36 135 577 657]{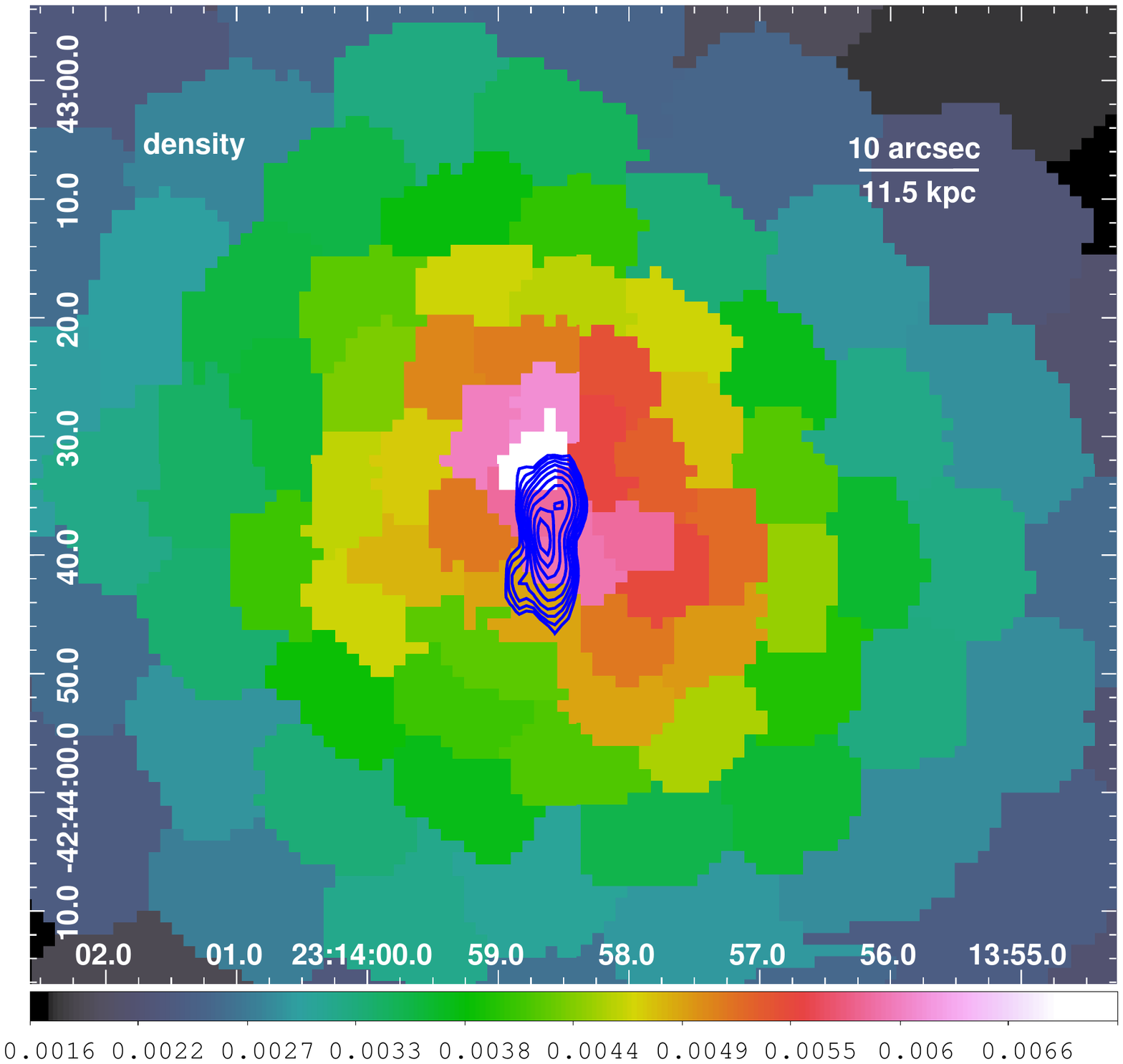}
\end{minipage}
\begin{minipage}{0.47\textwidth}
\includegraphics[height=8.3cm,clip=t,angle=0.,bb= 36 135 577 657]{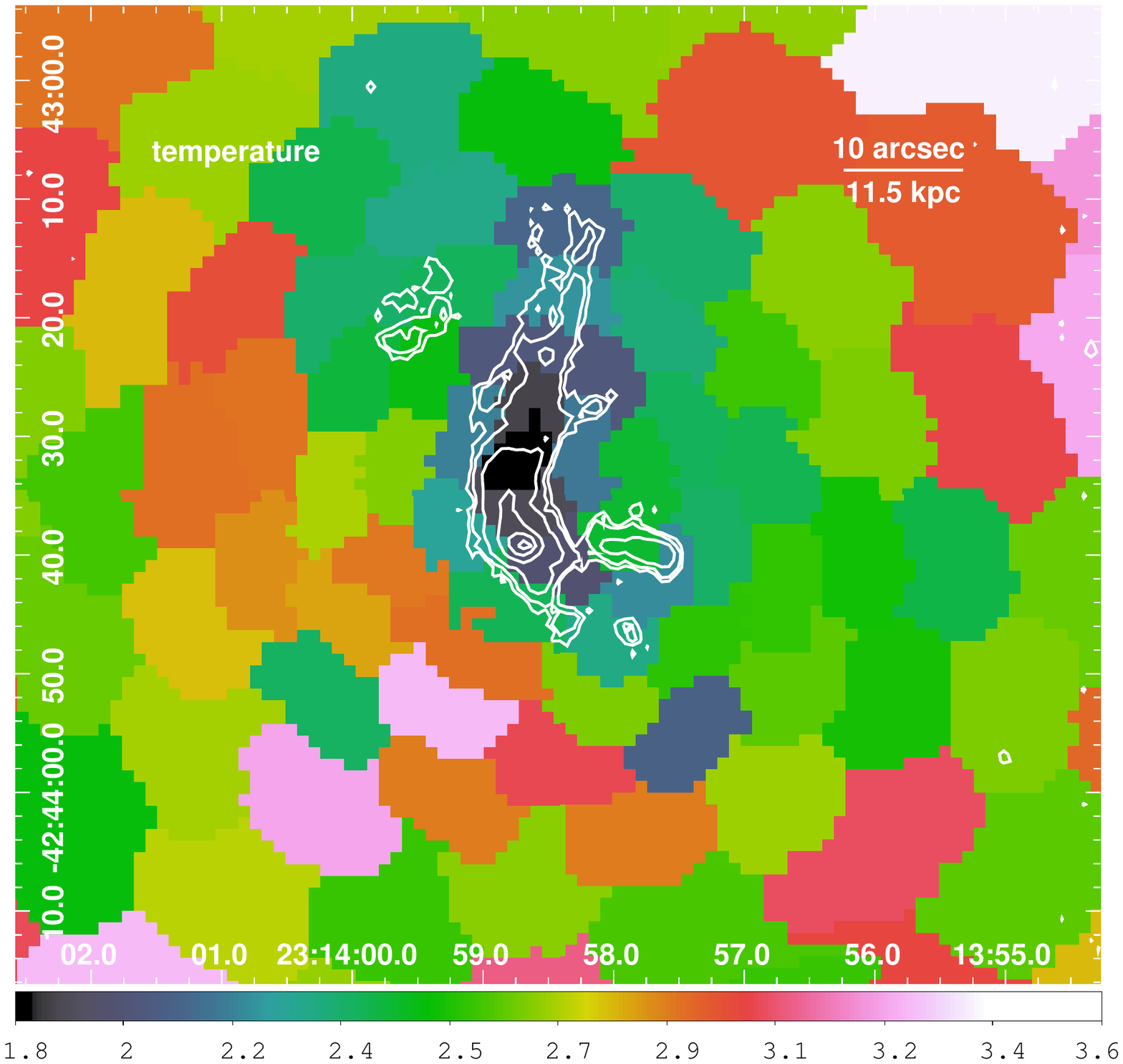}
\end{minipage}\\
\vspace{2mm}

\begin{minipage}{0.47\textwidth}
\includegraphics[height=8.3cm,clip=t,angle=0.,bb= 36 135 577 657]{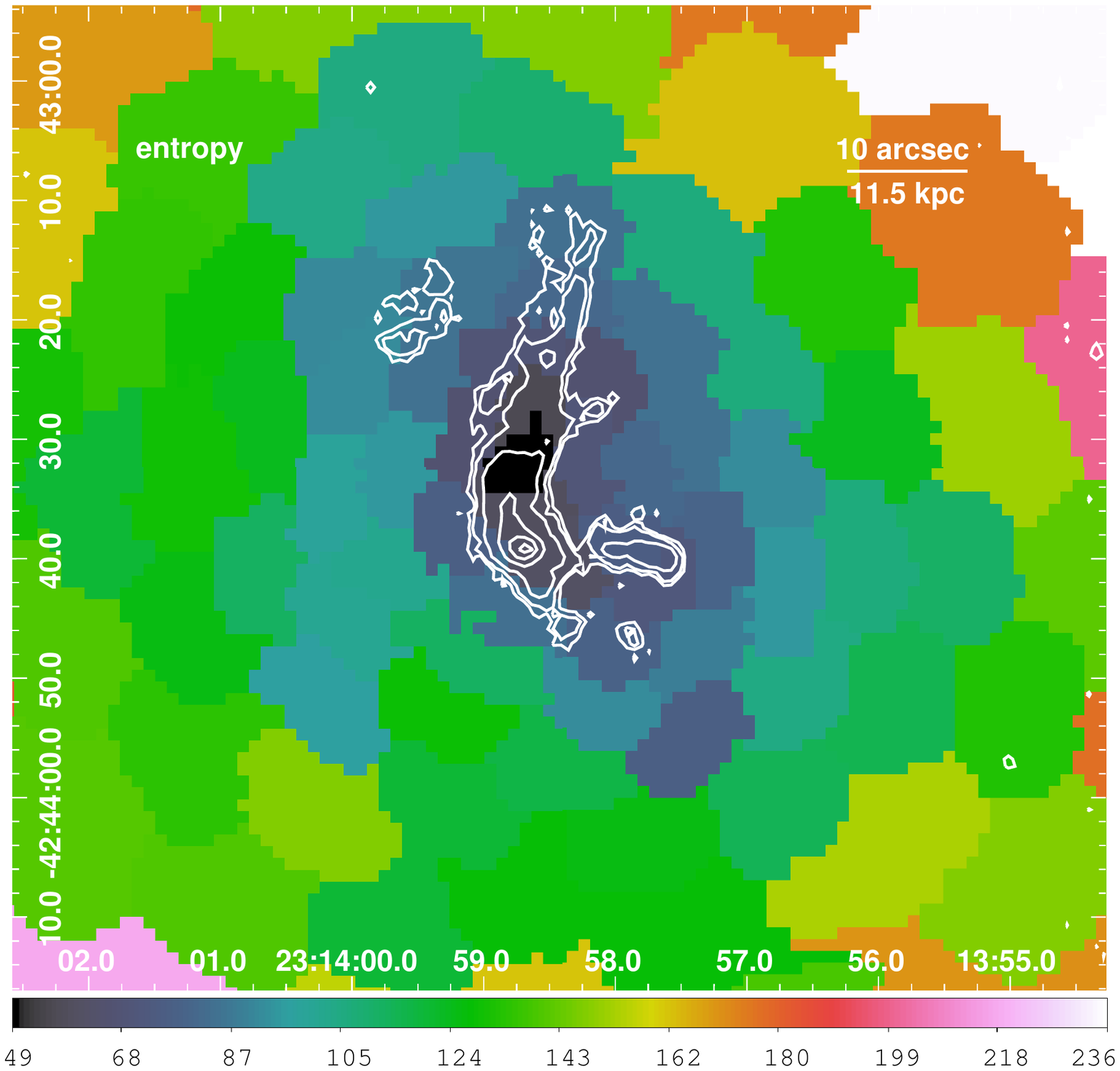}
\end{minipage}
\begin{minipage}{0.47\textwidth}
\includegraphics[height=8.3cm,clip=t,angle=0.,bb= 36 135 577 657]{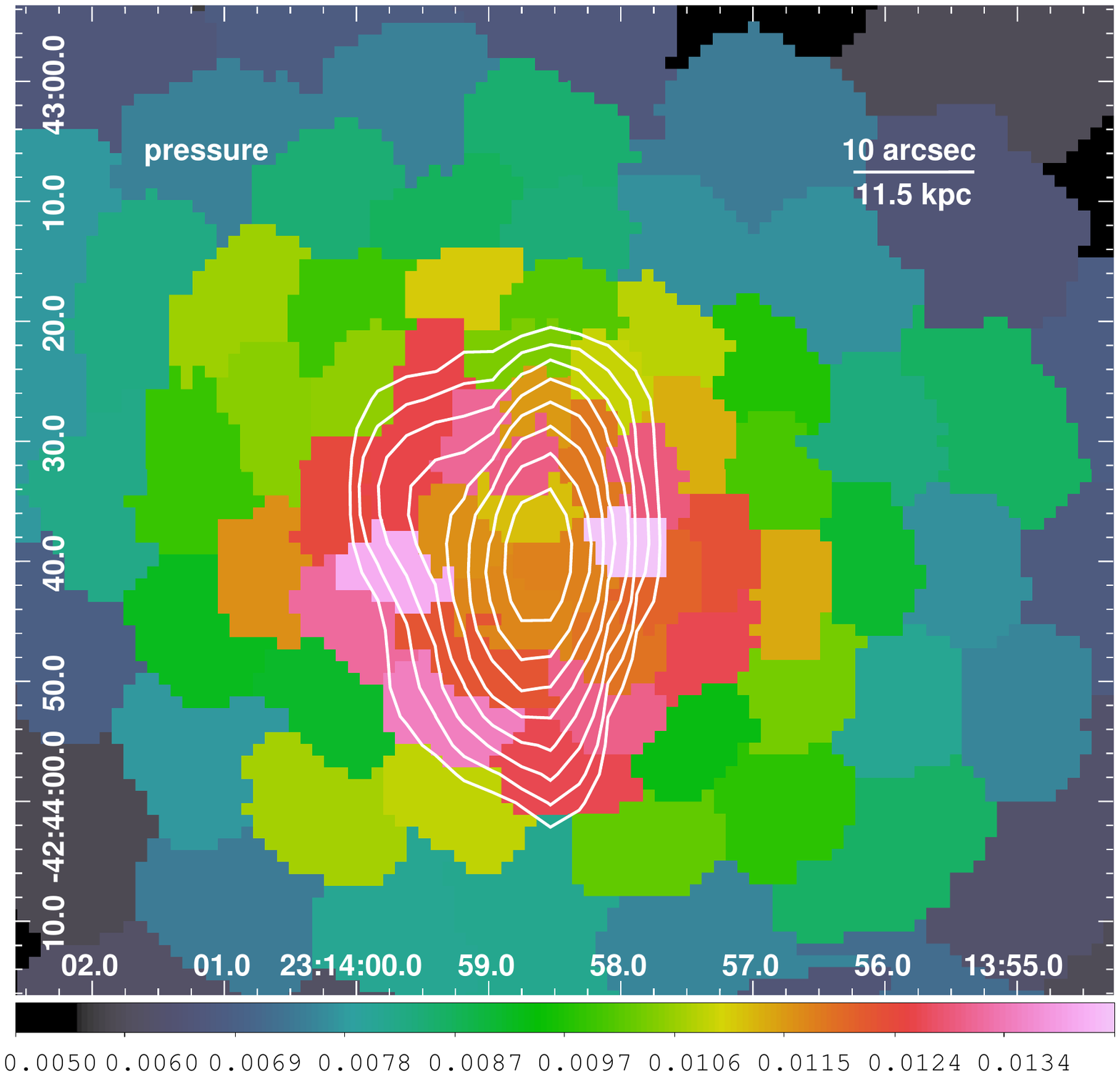}
\end{minipage}
\caption{2D maps of density (in units of cm$^{-3}$; top left panel), temperature (in units of keV; top right panel), pressure (in units of keV~cm$^{-3}\times\left(\frac{l}{\mathrm{2Mpc}}\right)^{-1/2}$; lower right panel), and entropy (in units of keV~cm$^2\times\left(\frac{l}{\mathrm{2Mpc}}\right)^{1/3}$; lower left panel). The maps were obtained by fitting each S/N$\sim$35 region independently with a single temperature thermal model, yielding 1$\sigma$ fractional uncertainties of $\sim$10 per cent on the temperature. The contours of the H$\alpha$+[\ion{N}{ii}] optical line emission are over-plotted  on the temperature and entropy maps. The  contours of the 8.4 GHz and 1.4 GHz radio emission are over-plotted on the density and pressure map, respectively. }
\label{thermo}
\end{figure*}

The X-ray data show no evidence for a point source associated with the central radio source. Assuming a power-law like spectrum with a photon index $\Gamma=2.0$, a conservative upper limit on the X-ray flux in the 2--10 keV band is $f_{\mathrm{X}}=2.3\times10^{-15}$~erg\, s$^{-1}$\, cm$^{-2}$. The large scale X-ray emission from the cluster is elongated in the same northeast-southwest direction as the major axis of the cD galaxy. The X-ray emission does not show obvious sharp surface brightness discontinuities, indicative of cold fronts due to sloshing gas. 

While the large scale X-ray morphology of S\'ersic~159-03 is relatively relaxed, its core is strongly disturbed. The brightest, densest X-ray emitting gas is displaced northward from the centre of the cD galaxy to a radius of $r\sim8$~kpc. In the top left panel of Fig.~\ref{images}, we show the smoothed \chandra\  image of the cluster core in the 0.5--7.5~keV band. The black cross indicates the position of the central AGN. In the top right panel, we show the same image, with the contours of the H$\alpha$+[\ion{N}{ii}] emission over-plotted in black, and the contours of the 8.4~GHz and 1.4~GHz radio emission over-plotted in blue and white, respectively. The lower panels of Fig.~\ref{images} show the same \chandra\ images divided by their best fit 2D elliptical double beta model. The images shows a bright ridge of dense thermal X-ray gas displaced by about 8~kpc to the north of the AGN and a clumpy X-ray filament extending along the H$\alpha$ filament to 31~kpc beyond this ridge. Another X-ray filament extends to the west and coincides with the western optical emission line nebula. 

The ridge of dense, thermal X-ray emitting gas to the north of the AGN seems to be interacting with, and confining, the 8.4~GHz radio emitting plasma (blue contours in Fig.~\ref{images}). The jets appear distorted by the interaction with the dense cooling gas. The 1.4~GHz radio plasma also appears deflected by the ridge of dense gas to the east, where it fills a gap - an elongated cavity - in the X-ray surface brightness distribution. This elongated cavity is possibly composed of two cavities. The X-ray surface brightness drops sharply to the southeast of the AGN, forming an apparent cavity with a radius of $\sim$11~kpc, filled by 1.4~GHz radio emission. This drop in surface brightness is spatially coincident with the sharp southeastern edge of the bright emission line nebulae.  A possible ghost cavity with a radius of $\sim$10~kpc, with no associated radio emission, can be seen about 30~kpc to the east of the nucleus.

\subsection{Thermodynamic properties of the core}
\label{sect:thermo}

The disturbed morphology of the cluster core is also reflected in the 2D maps of density, temperature, entropy, and pressure shown in Fig.~\ref{thermo}. The ICM spatially associated with the large north-south H$\alpha$ filament has a relatively low projected temperature (top right panel of Fig.~\ref{thermo}). The high density (top left panel of Fig.~\ref{thermo}) and low temperature of this feature translate into low entropy (lower left panel of Fig.~\ref{thermo}). The western filament, on the other hand, has a significantly higher temperature and entropy. 

The projected thermal pressure peaks in an approximate ring surrounding the core (lower right panel of Fig.~\ref{thermo}). The pressure in this ring is approximately 20 per cent higher than in the center. The maps reveal that the relatively over-pressured gas to the southeast of the core has a higher temperature than the surrounding gas, which strongly indicates that it has been compressed and heated by an AGN driven shock.

\begin{figure}
\includegraphics[width=0.95\columnwidth,clip=t,angle=0.]{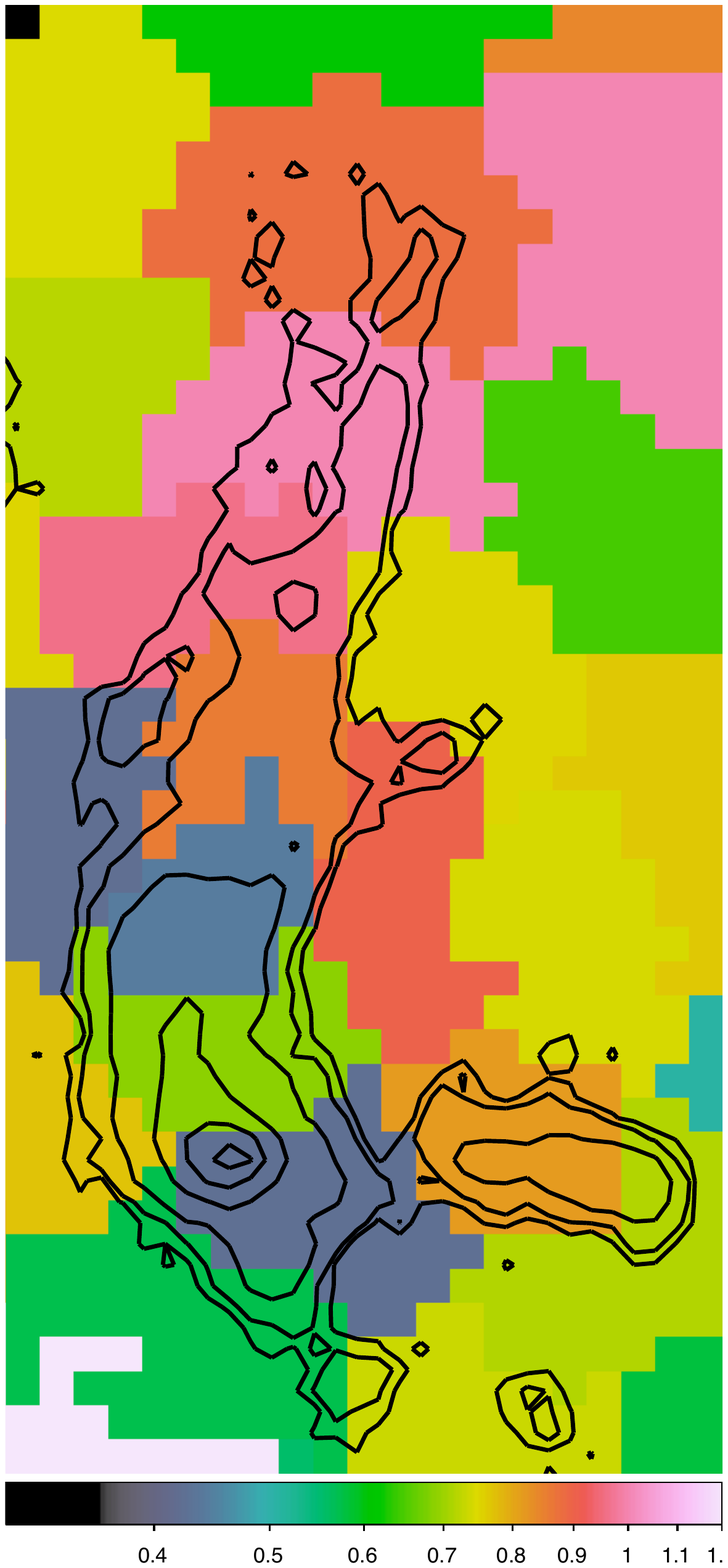}
\vspace{-0.5cm}
\caption{Metallicity map of the core of the cluster with the contours of the H$\alpha$+[\ion{N}{ii}] optical line emission over-plotted. The metallicity of the ICM along the filament is higher than that of the surrounding medium.  The apparent metallicity of the bright `northern ridge' and of the core is low, indicating the presence of multi-phase gas.  }
\label{metallicity}
\end{figure}

The metallicity of the ICM along the H$\alpha$+[\ion{N}{ii}] filaments is higher than that of the surrounding plasma (see Fig.~\ref{metallicity}). The apparent metallicities of the bright northern ridge of cooling plasma and of the core are, however, low. It has been shown that the metallicity is sensitive to the modeling of the underlying temperature structure and, if multi-temperature plasma is modeled with a single-temperature spectral model, the metallicity of gas can be significantly underestimated \citep[e.g.][]{buote2000b}.  The low metallicities of these features thus indicate the presence of multi-phase gas.

\subsection{High resolution X-ray spectra}

\begin{table}
\begin{center}
\caption{Best fit parameters for a four-temperature fit and a CIE+two-cooling-flow model fit to the high-resolution RGS spectra extracted from a 4\arcmin\ wide region centred on the core of S\'ersic~159-03. Emission measures, 
$Y=\int n_{\mathrm{H}}n_{\mathrm{e}}\mathrm{d}V$, are given in 10$^{66}$~cm$^{-3}$. Radiative cooling rates are given in $M_{\odot}$~yr$^{-1}$. The scale factor $s$ is the ratio of the observed LSF to the expected LSF based on the overall radial surface brightness profile. The upper limits are quoted at their 95 per cent confidence level. Abundances are quoted with respect to the values of \citet{grevesse1998}. }
\begin{tabular}{lcccccc}
\hline
Parameter			&  4T-model  		&   CIE+c.f.+c.f. model \\
\hline
$Y_{\mathrm{0.25keV}}$	&  $0.05\pm0.04$  		& --	\\
$Y_{\mathrm{0.75keV}}$	&  $0.19\pm0.03$	& -- 	\\
$Y_{\mathrm{1.5keV}}$	&  $0.7\pm0.3$		& --	\\
$Y_{\mathrm{3.0keV}}$	&  $12.1\pm0.3$ 	&   --    \\
$Y_{CIE}$			&	--			& $12.3\pm0.2$ \\
$\dot{M}_{\rm1.9-0.5keV}$	&	--			& $82\pm11$ \\
$\dot{M}_{\rm 0.5-0.1keV}$	&	--			& $<25$ \\
kT (keV)				&       --			& $3.05\pm0.16$	\\
s					& $1.12\pm0.13$	& $1.13\pm0.11$	\\
O					& $0.39\pm0.04$	& $0.43\pm0.04$	\\
Ne					& $0.43\pm0.12$	& $0.44\pm0.11$	\\
Fe					& $0.78\pm0.04$	& $0.82\pm0.06$	\\
\hline
\label{RGStable}
\end{tabular}
\end{center}
\end{table}

\begin{figure}
\includegraphics[width=1.1\columnwidth,clip=t,angle=0.]{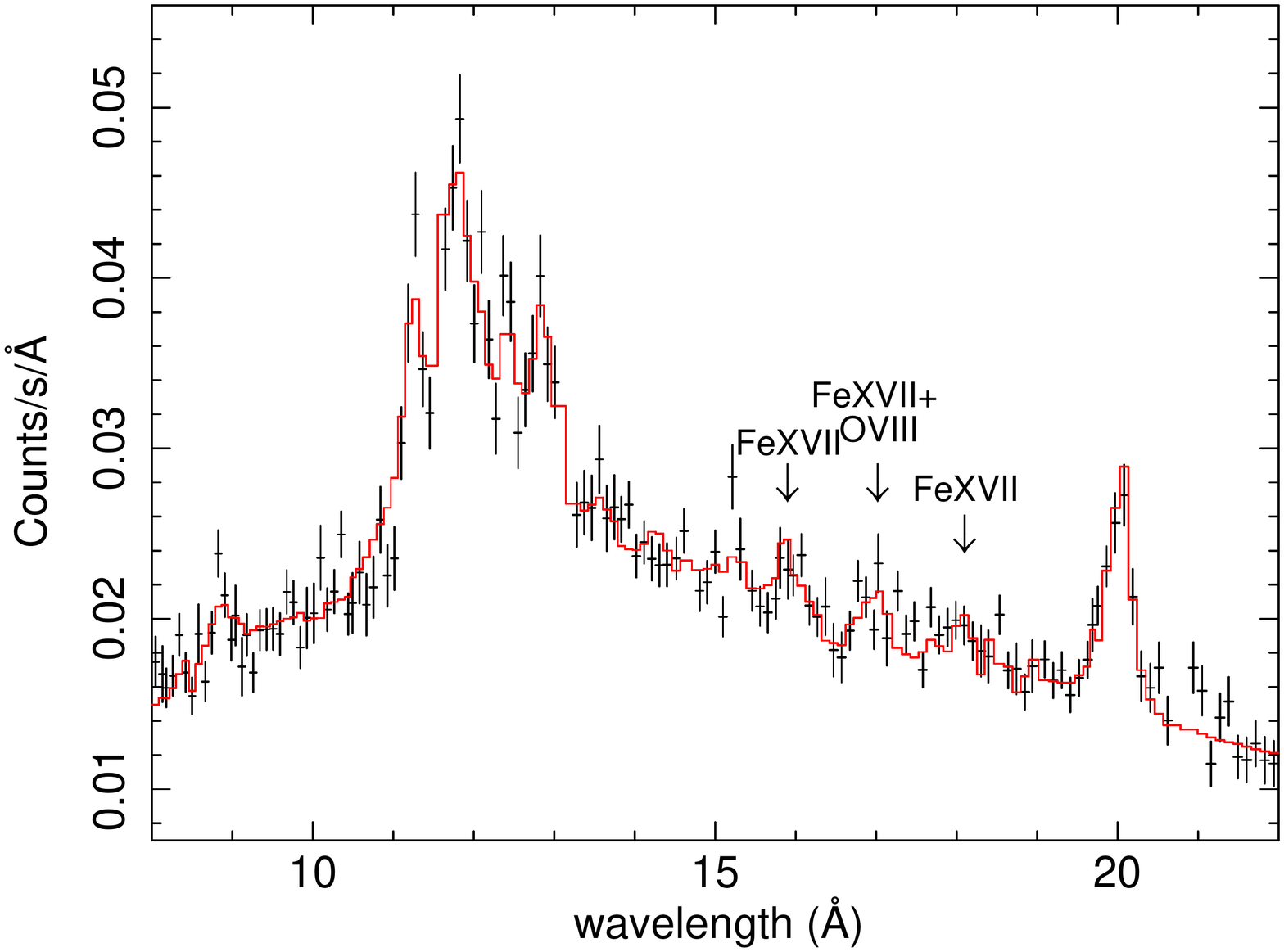}
\caption{The first order \xmm\ RGS spectrum extracted from a 4\arcmin\ wide region centred on the core of S\'ersic~159-03. The continuous line represents the best fit model to the spectrum. \ion{Fe}{xvii} lines emitted by plasma with $kT<0.9$~keV are clearly visible in the spectrum.  } 
\label{fig:rgs}
\end{figure}

In order to determine whether cooling X-ray gas with $kT<1$~keV is present in the core of the cluster, we examine the \xmm\ RGS spectra (see Fig.~\ref{fig:rgs}) to search for low temperature line emission (primarily \ion{Fe}{xvii}) tracing rapidly cooling gas. We fit the RGS data with a model consisting of collisionally ionized equilibrium plasmas at four fixed temperatures (0.25~keV, 0.75~keV, 1.5~keV, and 3~keV) with variable normalizations and common metal abundances. The abundances of O, Ne, Mg, and Fe are free parameters in the fit. Our fit confirms the presence of relatively cool, $kT<1$~keV plasma. Based on the lack of \ion{O}{vii} line emission however, which is emitted at $kT<0.4$~keV, we place a strong upper limit on the amount of gas cooling radiatively to very low temperatures. Our 90\% confidence upper limit on the emission measure of gas with $kT=0.25$ keV is $Y=\int n_{\mathrm{H}}n_{\mathrm{e}}dV=9\times10^{64}$ cm$^{-3}$, which is about 50\% of the best fit emission measure for the 0.75~keV gas. 

In order to place constraints on the amount of cooling in the cluster core we fit the RGS spectra with a separate model consisting of thermal plasma in collisional ionization equilibrium and two isobaric cooling flow models. The first cooling flow model is used to account for gas with  temperatures between $kT_\mathrm{upper}=1.9$~keV and $kT_\mathrm{Lower}=0.5$~keV, which appears to be the `temperature floor' in several cooling core clusters \citep{sanders2009,sanders2009b,werner2010}. The second cooling flow model is used to place an upper limit on the radiative cooling rate from 0.5 keV down to cold gas. While in the temperature range of 1.9--0.5~keV the data are formally consistent with a radiative cooling rate of $\dot{M}=82\pm11~M_{\odot}$~yr$^{-1}$, the 95\% confidence upper limit for radiative cooling from 0.5~keV to lower temperatures is only 25~$M_{\odot}$~yr$^{-1}$.  

\section{Discussion}

\subsection{Displacement of gas from the cD galaxy}

The core of S\'ersic~159-03 has a remarkably complex and rich morphology at all wavelengths. It displays signs of a powerful AGN feedback. The central regions of the galaxy ($r<7.5$~kpc) are cleared of the densest, X-ray emitting ICM and the cluster core displays a massive, bright H$\alpha$ filament extending northward from the centre of the cD galaxy to a radius of 35~kpc. This long filament is reminiscent of that in Abell~1795 \citep{fabian2001,crawford2005,mcdonald2009}. 

While the densest cooling X-ray plasma has been pushed away and uplifted from the cooling core by the radio jets to a radius of at least $r\sim8$~kpc, the brightest core of the H$\alpha$ emission is not displaced from the galactic nucleus. The most likely reason is that the H$\alpha$ emitting gas is only a thin layer on an underlying large reservoir of dense atomic and molecular gas at the base of the galaxy potential, which is difficult to uplift entirely. Observations of K-band emission lines of molecular and ionized hydrogen with the SINFONI integral field spectrograph reveal extended filaments in the core of S\'ersic~159-03, tracing closely each other and the H$\alpha$ emission \citep{oonk2010}. The line-of-sight velocities, as measured by \citet{oonk2010}, show a striking east/west dichotomy, suggesting that part of the velocity vector of the jets, which push and uplift the gas, is oriented along our line of sight. The gas to the east of the stellar core has a radial velocity of $\sim$200~km~s$^{-1}$ with respect to the cD galaxy, which seems to be decreasing with the distance from the nucleus.

The radial velocity distribution of the gas along the western filament goes smoothly from about -200~km~s$^{-1}$ to 50~km~s$^{-1}$ with increasing distance from the core. These velocity gradients indicate that the gas has been pushed out by the jets of the AGN. 

The morphology of the cooling filamentary gas indicates relative gas motions with the ambient ICM moving toward the northwest, dragging and bending the H$\alpha$ filament. The gas uplifted by the jet in the southwestern direction has been dragged to the west and the gas uplifted towards northeast has been displaced in the north-northwestern direction. These relative gas motions are most likely the result of the south-southeastward peculiar motion of the cD galaxy. 

Relatively cool X-ray gas is present along the whole northern filament of S\'ersic~159-03, but the correlation between the H$\alpha$ and soft X-ray emission is not perfect. We see clumps of dense, cooling X-ray emitting plasma at $\sim$8~kpc and at $\sim$17~kpc, but no obvious H$\alpha$ peaks at the same locations.  The `northern ridge' at $r\sim8$~kpc has the coldest projected temperature and lowest entropy. The X-ray filament extends to the north by about 3~kpc further than the observed H$\alpha$ filament. The western H$\alpha$+[\ion{N}{ii}] filament is also surrounded by bright, dense X-ray emitting gas, but its projected temperature is $\sim$0.8~keV higher. 
The relatively high metallicity of the filament compared to the ambient plasma (see Fig.~\ref{metallicity}) indicates that the filamentary system has been uplifted and stripped from the cD galaxy.

Gas uplift by buoyant radio-emitting plasma, supplied by the jets of the AGN, has also been observed in other cooling core clusters.
Detailed spectroscopic mapping of the cooling core of the Virgo Cluster, centered on M87, shows clearly that radio mode AGN feedback is highly efficient in stripping the core of the galaxy of its lowest entropy gas \citep{werner2010}. The total mass of uplifted gas in M87 is 6--9$\times10^8$~$M_{\odot}$, which is similar to the current gas mass within its innermost $r\sim3.8$~kpc region. The disruption of the core of S\'ersic~159-03 is considerably larger than that of M87, but by no means extreme. The AGN feedback in the Hydra A cluster is responsible for the uplift of a few times $10^9$ $M_{\odot}$ of low entropy plasma \citep{simionescu2009a}. Recently, \citet{ehlert2010} presented a multi-wavelength study of the cluster MACS~J1931.8-2634 where extreme AGN feedback with a jet power of $P_{\mathrm{jet}}\sim10^{46}$ erg~s$^{-1}$ and the sloshing gas disrupted the core, separating it into two X-ray bright ridges, which are currently at a distance of $\sim$25~kpc from the core. 

\subsection{Cooling of the displaced gas and star-formation}

To the north of the AGN, we see cooling X-ray plasma displaced by about 8~kpc from the nucleus, producing a prominent bright ridge which confines the high frequency radio plasma. The northern X-ray filament extends 31~kpc beyond this ridge. The relatively dense uplifted X-ray emitting gas, which is removed from the direct influence of the AGN jets, will cool in the absence of heating and eventually form stars. Multiphase cooling X-ray gas, displaced from the center of the cD galaxy and spatially coincident with H$\alpha$ emission, has also been seen in the Ophiuchus Cluster, in Abell 2052, and in MACS~J1931.8-2634 \citep{million2010,edwards2009,deplaa2010,ehlert2010}. 

Assuming that the filament has been uplifted at half of the sound speed (at 400~km~s$^{-1}$), the age of the filament will be $\sim10^8$~yr. That is approximately equal to the cooling time of 1~keV plasma in pressure equilibrium with the ambient ICM at the observed location. The \xmm\ RGS spectra clearly reveal \ion{Fe}{xvii} line emission associated with gas cooling to $kT<1$~keV. The 95~per cent confidence upper limit on radiative cooling below 0.5~keV is 25~$M_{\odot}$~yr$^{-1}$. 

The densest and coolest X-ray emitting clumps, in particular the bright multiphase `northern ridge', will cool and form narrow H$\alpha$ emitting filaments \citep[e.g. see simulations by][]{sharma2010}. It is possible that a significant fraction of the H$\alpha$ emitting gas in the northern filament at large radii is due to the cooling of the uplifted X-ray emitting plasma. The cold gas is likely to go on to eventually form stars. The \xmm\ OM UVW2 and \galex\ images \citep{mcdonald2010} indicate that the UV emission is extended along the brightest regions of the northern H$\alpha$ filament, indicating ongoing star formation.

The HST images show that the brightest regions of the filament are dusty. An infrared survey with the {\it Spitzer Space Telescope}, however, did not detect the system at 70$\mu$m \citep{quillen2008} indicating that the filaments do not contain an exceptionally large amount of warm dust. The excess UV emission from the cD galaxy corresponds to a star-formation rate of $\sim$2.3~$M_\odot$~yr$^{-1}$ for the assumed Salpeter IMF. Assuming the \citet{kennicutt1998} relation, SFR $(M_{\odot}$~yr$^{-1})=7.9\times10^{-42}$ ($L_{\mathrm{H}\alpha}$/erg~s$^{-1}$), and neglecting intrinsic extinction due to dust, the measured H$\alpha$ luminosity corresponds to a star-formation rate of 1.5~$M_{\odot}$~yr$^{-1}$. Accounting for internal extinction would further increase the intrinsic H$\alpha$ and NUV fluxes. Star-formation, however, is most likely not the only heating and ionizing source in the cluster center.

The large turbulent velocities and the elevated [\ion{N}{ii}]/H$\alpha$ ratio of $\sim$1.5 in the nucleus \citep{crawford1992} indicates that the interaction with the radio jets contributes strongly to the heating of the gas in the center. The [\ion{N}{ii}]/H$\alpha$ line ratios of 0.6 in the western filament are relatively high as well \citep{crawford1992}, indicating the presence of an additional non-ionizing source of energy at larger radii \citep[see e.g.][]{ferland2009}.

Based on the data for M87, \citet{werner2010} proposed that the H$\alpha$ filaments in its core may be powered by shock induced mixing of cold gas with the surrounding ICM \citep[see][]{begelman1990}. By bringing the hot thermal particles into contact with the cool gas, mixing can supply the power and ionizing particles to explain the observed spectra. Hot ICM electrons that penetrate into the cold gas excite the molecular hydrogen and deposit heat.   This scenario has been explored theoretically by \citet{ferland2008,ferland2009}, who studied heating by cosmic-rays, which affect the cooler ionized and neutral components in a similar way to hot ICM electrons.
The fact that the soft X-ray emission traces the H$\alpha$ emitting gas suggests that this process could be responsible for part of the H$\alpha$+[\ion{N}{ii}] line emission, part of the UV emission, and for the non-radiative cooling of the coldest X-ray gas in S\'ersic~159-03.

The most similar known X-ray/H$\alpha$ filament system is observed in Abell~1795. In that cluster the filament extends for 50~kpc.  The cD galaxy at the head of this filament is moving with respect to the ICM and the emission line nebula may originate from a runaway cooling of the hot X-ray emitting gas in the wake of that motion \citep{fabian2001,markevitch2001,mcdonald2009}. The filamentary system in S\'ersic~159-03, however, is most likely the result of AGN driven uplift and ram pressure stripping from the cD galaxy. 
The filament in Abell~1795 is composed of a pair of thin $w<1$~kpc intertwined H$\alpha$+[\ion{N}{ii}] filaments \citep{mcdonald2009}, spatially coincident with relatively cool X-ray emitting gas \citep{fabian2001}, and with chains of FUV-bright young star clusters condensing from the cooling gas in the filament \citep{crawford2005,mcdonald2009}.  
The northern end of the large H$\alpha$ filament in S\'ersic~159-03 separates into two narrow, parallel structures. Given a better spatial resolution, we would most likely resolve the filament into more thin threads. As discussed by \citet{fabian2008} for the emission line nebulae in the core of the Perseus Cluster, the thin thread-like filamentary structures are most likely stabilized by magnetic fields. These magnetic fields might be possible to detect using Faraday Rotation against the polarized jet emission \citep[as has been done for the filaments in the Centaurus Cluster by][]{taylor2007}.

\subsection{The powerful radio mode AGN}

The core of S\'ersic~159-03 harbors a powerful radio mode AGN. Inflating the southern cavity required a $4pV$ work of about $7\times10^{58}$~ergs, indicating powerful jets. However, no optical or X-ray point source is presently seen at the position of the radio bright AGN. Using the `Black hole fundamental plane' relation of \citet{merloni2003}, for a 10~mJy core flux at 4.75~GHz and for a black hole mass of $6\times10^8$~$M_{\odot}$ \citep[determined from the K-band bulge luminosity by][]{rafferty2006}, the expected X-ray core flux is $2.4\times10^{-13}$ erg\, s$^{-1}$\, cm$^{-2}$. This value is two orders of magnitude higher than our conservative upper limit of $2.3\times10^{-15}$~erg\, s$^{-1}$\, cm$^{-2}$. The observed scatter around the `Black hole fundamental plane' is, however, large. The radio luminosities of some other well known brightest cluster galaxies (e.g. NGC~1275) also show similar offsets with respect to the expected relation.   

The radio morphology is similar to 4C26.42 in Abell~1795 \citep{liuzzo2009} and it is also reminiscent of PKS~1246-410 in the Centaurus Cluster \citep{taylor2007}.  The distortion of the jets, and the change in the axis from N-S to E-W, is most likely due to the strong interactions with the dense gas. The velocity dispersion of the NIR emission lines, which sharply increases in the nuclear region \citep{oonk2010} shows that the interaction of the jets with the cold gas is the strongest within the innermost 2~kpc region.  The presence of cold, high density material in the cluster core is likely to decelerate the jets, which might entrain thermal gas, slow down to subsonic velocities, and continue to rise buoyantly. Such deceleration due to strong interaction with dense gas on small scales has been seen using VLBA observations in Hydra A \citep{taylor1996} and Abell~1795 \citep{liuzzo2009}. While the 1.4~GHz and 617~MHz radio plasmas appear to be deflected by the dense `northern ridge of cooling plasma' to the east, forming the `eastern elongated X-ray dark cavity', the younger radio jet seen at 8.4~GHz is being deflected to the northwest where it is likely to continue to buoyantly rise along the short axis of the cluster.  The 8.4~GHz southern jet is being deflected by the H$\alpha$ emitting gas to the southeast where it is partly filling the `southern X-ray cavity'.

Between 244~MHz and 1.4~GHz the integrated radio emission in the cluster core has a remarkably steep power-law spectrum $S_{\nu} \propto \nu^{-\alpha}$ with index $\alpha=1.49$. At higher frequencies, however, the radio spectrum flattens to index $\alpha=0.71$.  At the 4.75~GHz {\it ATCA} radio map (Hogan et al. in prep.), the radio morphology is very similar to that at 8.4~GHz, indicating that the flatter part of the integrated radio emission originates in the $\sim$10~kpc scale inner lobes (see the contours in the left panel of Fig.~\ref{radiospec}) where active jets are injecting relativistic particles. At the lower frequencies the radio emission is more extended, indicating an older population of electrons. The spectral index map produced from the 617~MHz and 1.4~GHz radio data (central panel of Fig.~\ref{radiospec}) shows that while in the central regions $\alpha\sim1$, to the east, where the radio plasma seems to be filling the elongated cavity, the spectral index steepens to $\alpha>2$. The spectral index also steepens in the northwest, where the plasma fills a region with a relative deficit in X-ray surface brightness. This radio plasma is most likely buoyant and exerting work on the surrounding medium.

The steep integrated spectrum of the radio source is similar to radio mini-halos found within the cooling radius of some clusters \citep[e.g][]{ferrari2008}. The morphology of the extended radio emission, however, indicates that it is due to the plasma from the radio jets which have been decelerated, deflected and confined by the interaction with the dense gas. A steep radio spectral index has been found in other cooling core clusters \citep[e.g., PKS 0745-191, A2029, A4059, A2597][]{taylor1994,pollack2005} and taken as an indicator of confinement. 

Our thermodynamic maps show that the projected thermal pressure peaks in an approximate ring surrounding the AGN. The pressure in the ring is approximately 20 per~cent higher than in the center. The high pressure region to the south of the AGN is hotter than the plasma both outside and inside this feature, strongly suggesting that it has been shock heated. The observed projected temperature jump of $\Delta T\sim1.2$ suggests a shock with a Mach number of $M\sim1.5$. The complex X-ray morphology with a cavity inside the shock front unfortunately prevents a more precise modeling of the shock. The relatively modest, factor of 1.5, temperature drop in the core of this cluster \citep{sun2009} also points towards a relatively strong AGN feedback activity. 
AGN induced shocks, which may be the most significant channel for heating of the ICM near to the AGN, have also been observed around M87 in the center of the Virgo Cluster and in Hydra~A,  \citep{forman2005,million2010,nulsen2005,simionescu2009b}.

\section{Conclusions}

We performed a multi-wavelength study of the energetic interaction between the central active galactic nucleus (AGN), the intra-cluster medium, and the optical emission line nebula in the galaxy cluster S\'ersic 159-03. We conclude that:

\begin{itemize}

\item 
Powerful `radio mode' AGN feedback and possible ram pressure cleared the central region of the cD galaxy ($r<7.5$~kpc) of the cooling low entropy X-ray gas. 

\item
This low entropy, high metallicity, relatively cool X-ray gas lies along the bright, 44 kpc long H$\alpha$+[\ion{N}{ii}] filament extending from the centre of the cD galaxy to the north. 

\item 
As indicated by the observed dust lanes, molecular and ionized emission line nebulae, and the excess UV emission, part of this displaced gas, which is removed from the direct influence of the AGN, cools and forms stars.

\item 
The pressure map shows evidence for an AGN induced weak shock at a radius of $r=15$~kpc and the X-ray images reveal cavities indicating past powerful AGN activity with an energy of $\sim7\times10^{58}$~ergs. At low frequencies, the radio source has an unusually steep spectrum with $\alpha=1.5$ indicating aging and confinement by the ambient gas. 

\end{itemize}

\section*{Acknowledgments}
We thank Alastair Edge and Michael Hogan for providing the reduced 4.75~GHz {\it ATCA} radio data and for helpful discussions. We thank Seth Bruch and Emily Wang for their help with the SOAR observations. 
Support for this work was provided by the National Aeronautics and Space Administration through Chandra/Einstein Postdoctoral Fellowship Award Number PF8-90056 and PF9-00070 issued by the Chandra X-ray Observatory Center, which is operated by the Smithsonian Astrophysical Observatory for and on behalf of the National Aeronautics and Space Administration under contract NAS8-03060, and by the \chandra\ grants GO0-11019X and GO0-11139X.  This work was supported in part by the U.S. Department of Energy under contract number DE-AC02-76SF00515. The Very Large Array is part of the National Radio Astronomy Observatory.  The National Radio Astronomy Observatory is a facility of the National Science Foundation operated under a cooperative agreement by Associated Universities, Inc. We thank the staff of the \gmrt\ that made these observations possible. The SOAR Telescope is a joint project of Conselho Nacional des Pesquisas Cientficas e Tecnologicas CNPq-Brazil, The University of North Carolina Chapel Hill, Michigan State University, and the National Optical Astronomy Observatory.

\bibliographystyle{mnras}
\bibliography{clusters}

\end{document}